\documentclass[twocolumn,aps,nofootinbib,superscriptaddress,tightenlines,notitlepage,pra,longbibliography]{revtex4-2}
\usepackage[english]{babel}
\usepackage{amsmath} 				
\usepackage{amssymb} 				
\usepackage{amsthm}
\usepackage{array}       				
\usepackage{graphicx}  			
\usepackage{xcolor}       			 
\usepackage[utf8]{inputenc} 		
\usepackage{url}     
\usepackage{times}
\usepackage{stmaryrd}

\usepackage[normalem]{ulem}

\usepackage[colorlinks=true,linkcolor=teal,citecolor=teal,urlcolor=teal]{hyperref}
\usepackage{bbold}
\usepackage{bm}
\usepackage{bbm}
\usepackage{microtype}

\usepackage[shortlabels]{enumitem}
\usepackage{braket}
\usepackage{mathtools}

\usepackage{tikz}
\usetikzlibrary{positioning}

\usepackage{hyperref}

\DeclareMathOperator{\sign}{sign}

\newcommand{\ketbra}[2]{\ket{#1}\!\bra{#2}}
\newcommand{\abs}[1]{\lvert #1 \rvert}
\newcommand{\norm}[1]{\lVert #1 \rVert}

\newcommand{\Norm}[1]{\left\lVert #1 \right\rVert}
\newcommand{\rest}{\mathrm{\textbf{R}}}
\newcommand{\trace}[1]{\mathrm{tr}\!\left(#1\right)}

\newcommand{\id}{\mathbb{I}}
\newcommand{\ee}{\mathrm{e}}
\newcommand{\ii}{\mathrm{i}}

\begin{document}

\title{Short-time simulation of quantum dynamics by Pauli measurements}

    \author{\href{https://orcid.org/0000-0002-8706-1732}{Paul\ K.\ Faehrmann}}
    \email{paul.faehrmann@fu-berlin.de}
	\affiliation{Dahlem Center for Complex Quantum Systems, Freie Universität Berlin, 14195 Berlin, Germany}
    \affiliation{Institute for Integrated Circuits and Quantum Computing, Johannes Kepler University Linz, Austria}

    \author{\href{https://orcid.org/0000-0003-3033-1292}{Jens\ Eisert}}
	\affiliation{Dahlem Center for Complex Quantum Systems, Freie Universität Berlin, 14195 Berlin, Germany}
    \affiliation{Helmholtz-Zentrum Berlin f\"{u}r Materialien und Energie, Hahn-Meitner-Platz 1, 14109 Berlin, Germany}
    \affiliation{Fraunhofer Heinrich Hertz Institute, 10587 Berlin, Germany}

    \author{\href{https://orcid.org/0000-0002-0749-8126}{Mária\ Kieferová}}
    \affiliation{Centre for Quantum Software and Information, School of Computer Science, Faculty of Engineering \& Information Technology, University of Technology Sydney, 
    NSW 2007, Australia}
    
    \author{\href{https://orcid.org/0000-0002-8291-648X}{Richard\ Kueng}}
    \affiliation{Institute for Integrated Circuits and Quantum Computing, Johannes Kepler University Linz, Austria}

\begin{abstract}
Simulating the dynamics of complex quantum systems is a central application of quantum devices. Here, we propose leveraging the power of measurements to simulate short-time quantum dynamics of physically prepared quantum states in classical post-processing using a truncated Taylor series approach. While limited to short simulation times, our hybrid quantum-classical method is equipped with rigorous error bounds. It is extendable to estimate low-order Taylor approximations of smooth, time-dependent functions of tractable linear combinations of measurable operators. These insights can be made use of in the context of Hamiltonian learning and device verification, short-time imaginary time evolution, or the application of intractable operations to sub-universal quantum simulators in classical post-processing.
\end{abstract}

\maketitle

\section{Introduction}

Predicting the dynamic properties of an interesting quantum system constitutes one of the most anticipated tasks to be tackled by nascent quantum computers and simulators. 
Recent years have seen unprecedented advances in the size (qubit number) and accuracy (gate fidelities) of such quantum hardware platforms, be it for universal quantum devices ( e.g.~\cite{kimEvidenceUtilityQuantum2023,reichardtDemonstrationQuantumComputation2024,acharyaQuantumErrorCorrection2024,reichardtLogicalComputationDemonstrated2024}) or sub-universal quantum simulators ( e.g.~\cite{bluvsteinLogicalQuantumProcessor2024,joshiExploringLargescaleEntanglement2023}).
However, full-fledged \emph{quantum simulation} is still prohibitively expensive and thus remains elusive.
Furthermore, purely classical methods are limited since a universal, efficient classical algorithm for quantum simulation would imply that all quantum algorithms could be efficiently simulated classically (${\rm BPP}={\rm BQP}$). This is because quantum dynamics generated by local Hamiltonians is ${\rm BQP}$ complete and hence as computationally powerful as a full quantum computer when formalized as a decision problem \cite{Vollbrecht,DynamicalStructureFactors,BQPTimeEvolution}.
Hence, a dedicated focus on resource efficiency is crucial to making the most out of currently available quantum platforms, be it through hybrid quantum-classical approaches outsourcing parts of the computation to classical hardware, designing algorithms tailored to specific hardware and simulation tasks, or improving classical methods.
To this end, trade-offs between the required circuit depth and the number of required circuit evaluations are of key interest since the former is highly limited in current, noisy quantum devices.

Following a second line of thought, in recent years, the notion of randomized measurements has gained traction in a variety of specific flavors, following a ``measure now, compute functions of interest later'' mindset~\cite{huangPredictingManyProperties2020c,elbenRandomizedMeasurementToolbox2023,morrisSelectiveQuantumState2020,painiEstimatingExpectationValues2021}
that is essential to the efficient functioning of most hybrid quantum-classical algorithms that can be executed on near-term hardware.
Besides estimating the energy of some trial states necessary for the variational quantum eigensolver and most other variational quantum algorithms~\cite{peruzzoVariationalEigenvalueSolver2014,bhartiNoisyIntermediatescaleQuantum2022, cerezoVariationalQuantumAlgorithms2021}, such methods can also be used to estimate expectation values of observables with small Frobenius norms, such as overlaps with pure states~\cite{huangPredictingManyProperties2020c,bertoniShallowShadowsExpectation2024} or estimating energy gaps with no additional resources beyond time evolution and measurements~\cite{chanAlgorithmicShadowSpectroscopy2024}. 
These applications thus impressively showcase the fundamental power of quantum measurements and classical post-processing and highlight the need for dedicated readout schemes such as grouping strategies~\cite{wuOverlappedGroupingMeasurement2021}, graph optimization techniques~\cite{verteletskyiMeasurementOptimizationVariational2020}, or fully randomized or de-randomized schemes~\cite{huangPredictingManyProperties2020c,elbenRandomizedMeasurementToolbox2023,huangEfficientEstimationPauli2021, hadfieldMeasurementsQuantumHamiltonians2022}.
After all, the power of quantum measurement goes hand in hand with efficient measurement protocols.

Randomness has found even more applications in reducing resource costs of near-term and early fault-tolerant quantum devices as it can help reduce errors that arise for product formulas~\cite{campbellShorterGateSequences2017b,campbellRandomCompilerFast2019a,childsFasterQuantumSimulation2019,chenConcentrationRandomProduct2021,faehrmannRandomizingMultiproductFormulas2022,chakrabortyImplementingAnyLinear2024}.
Furthermore, in the context of algorithms that use what are called \emph{linear combinations of unitaries (LCU)}~\cite{lowWellconditionedMultiproductHamiltonian2019,berryHamiltonianSimulationNearly2015, berryExponentialImprovementPrecision2014, berrySimulatingHamiltonianDynamics2015b, lowHamiltonianSimulationQubitization2019b,faehrmannRandomizingMultiproductFormulas2022}, randomness can help circumvent the need for block encodings~\cite{faehrmannRandomizingMultiproductFormulas2022,wangQubitEfficientRandomizedQuantum2024,wanRandomizedQuantumAlgorithm2022}.
For certain settings, these LCUs can be constructed purely in post-processing~\cite{vazquezWellconditionedMultiproductFormulas2023} without actually requiring any auxiliary qubits.
\begin{figure*}[t!]
    \centering
    \includegraphics[width=.99\linewidth]{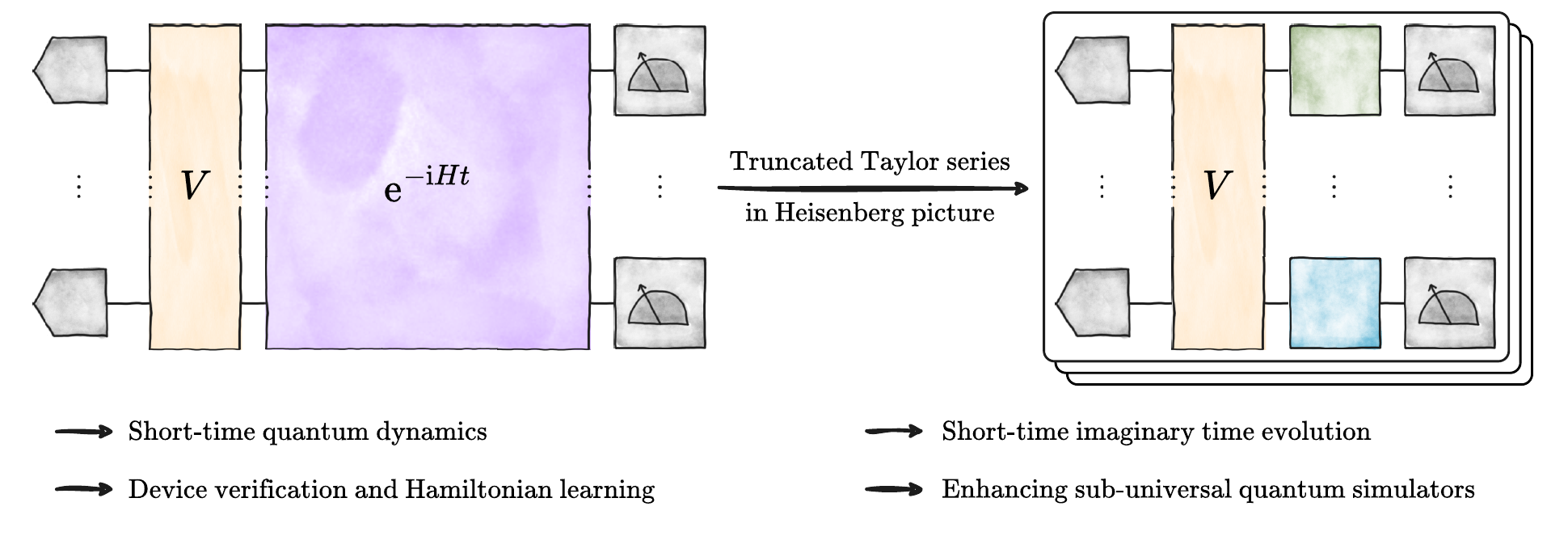}
    \caption{\emph{Cartoon illustration of our framework:} 
    We use a truncated Taylor series approach in the Heisenberg picture to achieve short-time quantum simulation entirely via classical post-processing. This trade-off between circuit depth (\emph{left}) and sample complexity (\emph{right}) only requires the state preparation unitary $V$ (potentially classically intractable Schrödinger evolution), as well as one parallel layer of `reprogrammable' single-qubit Clifford gates (think: local Clifford/Pauli shadows).
   }
    \label{fig:overview}
\end{figure*}
While these methods approach the task of making quantum simulation more feasible from the side of improving quantum algorithms or trading resources to accommodate noisy and early fault-tolerant quantum devices, there has also been a lot of recent progress toward improved purely classical methods simulating quantum dynamics, using tensor network methods~\cite{osborneEfficientApproximationDynamics2006,Daley,orusTensorNetworksComplex2019}, truncated Pauli paths~\cite{schusterPolynomialtimeClassicalAlgorithm2024} or cluster expansions~\cite{wildClassicalSimulationShortTime2023}.
Cluster expansion methods are well-suited for the classical simulation of short-time quantum dynamics for product states evolving under a few-body Hamiltonian.
They are reminiscent of the quantum algorithm simulating Hamiltonian dynamics with truncated Taylor series~\cite{berrySimulatingHamiltonianDynamics2015b}, another LCU approach, and use nested commutators to reduce computational costs.
Using backpropagated Heisenberg evolution has also sparked recent interest in the context of Pauli path methods for classically simulating quantum systems~\cite{shaoSimulatingNoisyVariational2024,begusicFastConvergedClassical2024,angrisaniClassicallyEstimatingObservables2024,angrisaniSimulatingQuantumCircuits2025a}.
These conceptual similarities raise the crucial question of whether such methods can be used in a hybrid quantum-classical setting to help enhance current quantum devices.

This work demonstrates how the power of randomized Pauli measurements can be leveraged to estimate short-time dynamics of arbitrary quantum states in classical post-processing, thereby trading circuit depth for additional measurements.
It is inspired by viewing the results for simulating Hamiltonian dynamics with a truncated Taylor series~\cite{berrySimulatingHamiltonianDynamics2015b} within the Heisenberg picture.
The proposed method is in line with other recent insights on the power of measurements for the manipulation of quantum states \cite{PhysRevX.11.011030,PhysRevX.9.031009} and strengthens the ``measure now, compute functions of interest later'' mindset~\cite{huangPredictingManyProperties2020c,elbenRandomizedMeasurementToolbox2023,morrisSelectiveQuantumState2020,painiEstimatingExpectationValues2021}.

Moreover, this framework readily extends beyond quantum dynamics and applies to any low-order Taylor approximation of spectral functions of linear combinations of measurable operators, not just Pauli operators. 
As such, we anticipate further practical applications in notions of \emph{Hamiltonian learning} that are based on short-time dynamics or in \emph{device verification}, in \emph{short-time imaginary time evolution}, or for introducing operations intractable to a given \emph{sub-universal quantum simulator} in classical post-processing. 
These applications will be sketched here.

We refer to Figure~\ref{fig:overview} for a cartoon illustration of this framework and its versatility.

\section{Methods}
\subsection{Problem setting} 

The precise problem we address in this work is the following: Consider an $n$-qubit system whose Hamiltonian $H$ is specified as the linear combination of $L$ Pauli operators $H_\ell$ acting on an $n$-qubit system, that is,
\begin{equation}
    H = \sum_{\ell=1}^{L}\alpha_\ell H_\ell,\quad \text{with }\quad\lambda=\sum_{\ell=1}^{L}|\alpha_\ell|,
\end{equation}
where the number of non-trivial tensor factors per Pauli operator, also denoted as their Pauli weight, is upper bounded by $w$. 
Furthermore, let $O$ be some (linear) observable. 
Then, the goal is to estimate the time dynamics $\trace{\ee^{\ii H t}O\ee^{-\ii H t}\rho}$ for an arbitrary (potentially mixed) initial state $\rho$ and (short) evolution time $t>0$.

In a more general sense, we are interested in estimating quantities of the form $\trace{ F(H,t)\rho}$, where $(H,t)\mapsto F(H,t)$ can be any function depending on a linear combination of Pauli operators $H$ and variable $t$ whose truncated Taylor series with respect to $t$ can be bounded. 
To avoid a convoluted presentation, we will focus on the specific application of estimating time dynamics and mention how the procedure would differ in a more general setting.

\subsection{Expanding time dynamics with a Taylor series}
\label{sec:expanding_dynamics}

Using the approach of employing a truncated Taylor series for Hamiltonian simulation proposed in Ref.~\cite{berrySimulatingHamiltonianDynamics2015b}, we can approximate such functions by implementing their Taylor series truncated at $K$-th order, i.e.,
\begin{align}
\nonumber
    U(t)&\coloneqq\ee^{-\ii H t}\approx\sum_{k=0}^K\frac{(-\ii Ht)^k}{k!}\\
    &\approx \sum_{k=0}^K\sum_{\ell_1,\ldots,\ell_k=1}^{L}\underbrace{\frac{(-\ii t)^k}{k!}\alpha_{\ell_1}\cdots\alpha_{\ell_k}}_{\beta_j}\underbrace{H_{\ell_1}\cdots H_{\ell_k}}_{P_j}\\
    \widetilde{U}(t)&=\sum_{j=1}^m \beta_j P_j \label{eq:U_expansion}
\end{align}
which itself is the sum of $m \leq\sum_{k=0}^K L^k=\left(L^{K+1}-1\right)/(L-1)=\mathcal{O}\left(L^K \right)$ $n$-qubit Pauli operators $P_j$ corresponding to some $H_{\ell_1}\cdots H_{\ell_k}$ (whose weight is upper bounded by $\min\{wK,n\}$) and associated pre-factors $\beta_j$.
As long as $\norm{H}t\leq \lambda t\leq1$, e.g., in the case of short times, the error of this approximation is upper-bounded by $\epsilon$ when choosing $K=O(\log(1/\epsilon))$. 
Appendix~\ref{A:tail_bound} thoroughly discusses and analyzes the truncation error of a Taylor approximation of the time evolution operator, recovering previous results of Ref.~\cite{berrySimulatingHamiltonianDynamics2015b}.
However, this is not the case anymore for longer simulation times. 
One way to circumvent this issue is to slice the time evolution into $r\geq\norm{H}t$ segments $U(t)=U(t/r)^r$, each with error $\epsilon/r$, increasing the number of terms for the total time evolution in Eq.~\eqref{eq:U_expansion} exponentially with respect to $r$ in worst-case.

Note that the time-evolved observable $F(H,t)=\ee^{\ii Ht}O\ee^{-\ii H t}=O(t)$ features two unitary evolutions. We can approximate both of them separately and obtain
\begin{align}
    \label{eq:o_concatenated}
    \widetilde{O}(t)&= \widetilde{U}(-t)O\widetilde{U}(t)=\sum_{j,j'=1}^m \underbrace{\beta_j \beta_{j'}}_{\gamma_i} \underbrace{P_jOP_{j'}}_{Q_i} \\
    &=\sum_{i=1}^{m^2}\gamma_iQ_i \label{eq:o_expansion},
\end{align}
which is now a sum of at most $m^2$ different $n$-qubit Pauli operators of the form $Q_i=P_jOP_{j'}=H_{\ell_1}\cdots H_{\ell_k}OH_{\ell_1}\cdots H_{\ell_{k'}}$.
A priori, the coefficients $\gamma_i=\beta_j\beta_{j'}$ can be complex-valued, but the underlying Hermitian structure of the time-evolved observable (i.e.\ symmetry under taking adjoints) ensures that complex-valued coefficients must cancel out and the total sum ranges over $m_{\mathrm{tot}} \leq m^2$ terms with real-valued coefficients $\gamma_i \in \mathbb{R}$.

Estimating $\trace{O(t)\rho}$ is thus well-approximated by a real-valued, linear combination of the expectation values of Pauli operators $Q_i$, i.e.,
\begin{equation}
    \label{eq:measurement_expansion}
    \trace{O(t)\rho}\approx\trace{\widetilde{O}(t)\rho}=\sum_{i=1}^{m_\mathrm{tot}}\gamma_i \trace{Q_i\rho},
\end{equation}
where $m_\mathrm{tot}$, the total number of Pauli operators to be estimated, can be significantly smaller than the upper bound of $m^2$ from Eq.~\eqref{eq:o_expansion}.

For $K=O(\log(r/\epsilon))$ and $r=\norm{H}t$, the number of Pauli expectation values in this sum is upper bounded by 
\begin{align}
    m_\mathrm{tot}&\leq \left(\sum_{k=0}^K L^k\right)^{2r}\!\!=\mathcal{O}(L^{2K r})\\
    &=\mathrm{poly}\left(\left(\frac{\norm{H}t L}{\epsilon}\right)^{\norm{H}t}\right),\label{eq:nr_of_terms}
\end{align}
where $\| H \|$ denotes the operator (or spectral) norm of the Hamiltonian $H$.
Here, it is important to note that, strictly speaking, the dependence on time is (weakly) superexponential, as is the dependence on the system size, which indirectly enters via $\norm{H}$.

The maximum Pauli weight over all $Q_i$s is given by 
\begin{equation}
    w_\mathrm{max} = \min(\mathcal{O}\left(2ww_O\norm{H}t\log(\norm{H}t)),n\right),
\end{equation}
where we additionally need to take the maximum Pauli weight $w_O$ of the Pauli expansion of the observable $O$ into account. 

Therefore, estimating time dynamics can effectively be reduced to a pure measurement task, albeit with potentially many different measurement configurations (single-qubit Pauli measurements). 
Since the number of expectation values of Pauli operators to be estimated grows exponentially in the simulation time, this tradeoff only pays off for short simulation times. 

Comparing these insights with the original algorithm for simulating quantum dynamics with a truncated Taylor series~\cite{berrySimulatingHamiltonianDynamics2015b}, we find a similar exponential scaling, although in quantum circuit depth instead of sample complexity as we observe here.
 
It requires the implementation of the linear combination of unitaries from Eq.~\eqref{eq:U_expansion}, which does contain quadratically fewer terms than Eq.~$\eqref{eq:o_expansion}$ but also requires LCU techniques that increase the circuit depth further~\cite{childsHamiltonianSimulationUsing2012b}.
The presented algorithm thus swaps a deep circuit size for the task of estimating many different Pauli expectation values of Eq.~\eqref{eq:measurement_expansion}.
The latter is much more feasible on current and near-term quantum hardware. 

It is important to note that since the Taylor expansions rely on powers of the Hamiltonian and, thus, powers of sums of Pauli operators, not all will be pairwise different. 
This is due to their group structure and the fact that Pauli operators square to the identity. 
Consequently, the effective number of terms in this sum can be significantly smaller than the provided upper bound. 
However, an exponential scaling in the chosen cutoff $K$ and number of segments $r$ cannot be fully avoided.

For time-evolved observables $O(t)$ in particular, using the well-known fact that the time evolution in the Heisenberg picture for time-independent Hamiltonians and observables results in well-behaved nested commutators resulting from the Baker–Campbell–Hausdorff formula (see e.g.~), we find that instead of concatenating two truncated Taylor series of the time evolution operator as in Eq.~\eqref{eq:o_concatenated}, we can use
\begin{equation}
    \label{eq:nested}
    \ee^{\ii H t}O\ee^{-\ii H t}=O + \ii t[H,O]+\frac{(\ii t)^2}{2!}[H,[H,O]]+\ldots,
\end{equation}
which is reminiscent of cluster expansion techniques \cite{wildClassicalSimulationShortTime2023} using nested commutators to decide which terms to estimate in a classical algorithm on product states and stressing that we are leveraging the group structure of the Pauli operators.
The advantage of such a reformulation is that we no longer have to calculate all combinations of Pauli operators to obtain the $Q_i$ in Eq.~\eqref{eq:o_expansion}, many of which will cancel, but can use the underlying structure to calculate only non-vanishing terms.
The connection to well-understood cluster expansion techniques provides a straightforward path to enhancing such techniques with quantum resources.
Instead of performing such an expansion and estimating the individual Pauli expectation values classically for product states as in Ref.~\cite{wildClassicalSimulationShortTime2023}, we can measure them on arbitrary quantum states using existing expansions.
Since nested commutators are a reformulation of the Taylor series of $O(t)$, the same error bounds as in Appendix~\ref{A:direct_expansion} apply.
Central to cluster expansion techniques and extensions thereof, thorough discussions can be found, e.g., in Refs.~\cite{wildClassicalSimulationShortTime2023,bakshiLearningQuantumHamiltonians2023}.
Knowledge of the structure of the function of interest can thus help reduce the amount of classical post-processing.

Instead of concatenating two Taylor approximations of $U(t)$ for $t>0$ into an approximation of $O(t)$, as we have presented above, we can also directly Taylor expand the target function, as one would do for more generic functions $F(H,t)$.
How such a direct expansion would present itself and differences in error and the number of terms are discussed in Appendix~\ref{A:direct_expansion}.

\subsection{Error bounds}
An approximation of $\trace{F(H,t)\rho}$ via a truncated Taylor series followed by an estimation of the resulting linear combination of expectation values of Pauli operators is accompanied by two types of errors: A systematic error due to the truncation of the Taylor series and a sampling error obtained during the estimation of the Pauli expectation values. 
Both of these error types can be rigorously bounded.

\subsubsection{Systematic errors}
\label{subsec:systematic_error}

We can upper-bound the operator norm difference between the truncated Taylor series and target matrix function $F(H,t)$  by employing Taylor's remainder theorem as showcased in Appendix~\ref{A:tail_bound} for a direct Taylor expansion of the particular matrix-valued function $F(H,t)=O(t)$.
When stitching together multiple approximations as in Eq.~\eqref{eq:o_concatenated} above, where we have constructed an approximation of $O(t)$ using a truncated Taylor series for $U(t)$, we can also combine the individual approximation errors to bound the total error when approximating the target function.
As in Ref.~\cite[Lemma 1]{faehrmannRandomizingMultiproductFormulas2022}, given an approximation $\widetilde{U}$ of $U$ with $\norm{U-\widetilde{U}}\leq \epsilon$, we can redefine $\widetilde{U}=U+U_\epsilon$ with $\norm{U_\epsilon}\leq\epsilon$.
We then find
\begin{align}
    \Norm{U^\dagger OU-\widetilde{U}^\dagger O\widetilde{U}}&=\Norm{U_\epsilon^\dagger OU +U^\dagger OU_\epsilon+U_\epsilon^\dagger O U_\epsilon}\\
    \nonumber
    &\leq 2 \norm{U}\norm{U_\epsilon}\norm{O}+\norm{U_\epsilon}^2\norm{U}\\
    \nonumber
    &\leq (2\epsilon+\epsilon^2)\norm{O}\\
    \nonumber
    &\leq 3\epsilon \norm{O},
    \nonumber
\end{align}
providing a rigorous error bound for the systematic error of the approximation $\widetilde{O}(t)$ defined in Eq.~\eqref{eq:o_concatenated} to the actual time dynamics.
Using techniques from Ref.~\cite{childsTheoryTrotterError2021}, the error bound could be further tightened using nested commutators.

These bounds constitute only the systematic error originating in the truncation of the Taylor series and not the error obtained in implementing the linear combinations of Pauli operators, which is method-dependent and would have to be added to bound the actual error picked up during implementation.

\subsubsection{Sampling error}

Using state-of-the-art methods, the linear combination of unitaries of Eq.~\eqref{eq:o_expansion} could be implemented deterministically using block encodings in the spirit of Refs.~\cite{childsHamiltonianSimulationUsing2012b,berrySimulatingHamiltonianDynamics2015b} or in a randomized fashion using the frameworks of Refs.~\cite{wanRandomizedQuantumAlgorithm2022,faehrmannRandomizingMultiproductFormulas2022}. 
However, since we are dealing with sums of Pauli operators, simply performing Pauli basis measurements as outlined in Eq.~\eqref{eq:measurement_expansion} suffices, and no further quantum circuitry is required.

Thus, estimating $\trace{\ee^{\ii H t}O\ee^{-\ii H t}\rho}$ is equal to estimating the expectation values of the sum of Pauli operators given by the nested commutator in Eq.~\eqref{eq:nested} for some initial state $\rho$.
Therefore, besides preparing the input state $\rho$, the main cost of this algorithm lies in the number of measurements required to obtain sufficient statistical data for all terms of the truncated Taylor series.

This sampling error can be approached with two different mindsets: we can view these methods as the primary objective of an experiment or as secondary to another task. 
That is, we either care only about $\trace{F(H,t)\rho}$ or are performing another experiment, such as a variational algorithm estimating $\trace{H\rho(\boldsymbol{\theta})}$ for some trial state $\rho(\boldsymbol{\theta})$, using some readout scheme that produces classical data from which one could, albeit sub-optimally, also infer information about $\trace{F(H,t)\rho}$.

\paragraph{Direct estimation}
If this method is the primary objective, we want to reduce the sampling error as much as possible while also being practical, i.e., reducing the classical pre-processing cost. 
To this end, we propose using a randomized measurement strategy, although any measurement scheme is viable.
The advantage of a randomized scheme is that each sample taken from the quantum devices will then be an unbiased estimator of the whole sum with corresponding convergence properties. 
To simplify notation, we denote the Taylor expansion as a linear combination of weighted Pauli operators, pulling the sign of each pre-factor into their corresponding Pauli, i.e.,
\begin{align}
\widetilde{O}(t)&=\sum_{i=1}^{m_\mathrm{tot}}\gamma_iQ_i=\sum_{i=1}^{m_\mathrm{tot}}\abs{\gamma_i} (\sign(\gamma_i)Q_i)\\
&=\norm{\gamma}_1\sum_{i=1}^{m_\mathrm{tot}} \underbrace{\frac{\abs{\gamma_i}}{\norm{\gamma_i}_1}}_{p_i}\underbrace{(\sign(\gamma_i)Q_i)}_{P_i},    
\nonumber
\end{align}
where we further normalize the pre-factors $\gamma_j$ such that we arrive at a probability distribution $\{p_i,P_i\}$. 

Since we are concatenating two Taylor approximations of the time evolution operator $U(t)=\ee^{\ii Ht}$, where the 1-norm of the terms of the truncated Taylor series is upper bounded by $\ee^{\lambda t}$, we have 
\begin{equation}
    \norm{\gamma}_1=\mathrm{poly}\left(\ee^{\lambda t}\right),
\end{equation}
directly influencing the number of required samples.
A tighter bound can be obtained by truncating the Taylor expansion of $\ee^{\lambda t}$ at the same order that the Taylor series of $\ee^{\ii Ht}$ is truncated.

Here, $\lambda$ will implicitly scale with the system size and thus the number of terms in the Hamiltonian, resulting in an exponential scaling with respect to the system size, which is likely to be a limiting factor in practice.

These insights are reminiscent of the runtime for cluster expansion methods~\cite[Theorem 6]{wildClassicalSimulationShortTime2023}.

Sampling which Pauli expectation value $\trace{P_i\rho}$ to measure in each circuit evaluation from this distribution results in an unbiased estimator of $\widetilde{O}(t)$.
Using Hoeffding's inequality, we then find that
\begin{equation}
    N\geq \frac{2\norm{\gamma}_1^2\ln(2/\delta)}{\epsilon^2}
\end{equation}
samples suffice to estimate the linear combination of Pauli operators up to an error of $\epsilon$ with probability at least $1-\delta$.
Note that just estimating $\trace{O\rho}$ for the same $\epsilon$ and $\delta$ would require $N\geq 2\ln(2/\delta)\epsilon^{-2}$ samples, meaning that this method increases the number of required samples by a factor of $\norm{\gamma}_1^2$.
The advantage of an importance sampling procedure is its independence on the number of terms $m_\mathrm{tot}$.
Further, an importance sampling, first of the Taylor order and then of the terms in the Hamiltonian, can avoid the need for classical preprocessing to construct an explicit expansion.

This increase in sample complexity and the consequential limit to short times is precisely what the quantum algorithm for a randomized truncated Taylor series presented in Ref.~\cite{wanRandomizedQuantumAlgorithm2022} avoids with quantum resources to obtain a controllable overhead of $\exp(\lambda^2t^2/r)$.

The estimation suggested here can be improved using structural knowledge of the target function, e.g., by incorporating the nested commutator structure and quantum resources saved by immediately returning zero for all complex-valued $\gamma_j$ when considering $O(t)$.
Alternatively, grouping strategies~\cite{wuOverlappedGroupingMeasurement2021}, graph optimization techniques~\cite{verteletskyiMeasurementOptimizationVariational2020}, fully randomized measurements~\cite{huangPredictingManyProperties2020c,elbenRandomizedMeasurementToolbox2023}, or other dedicated readout schemes could be used with their corresponding sampling error bounds.

\paragraph{Using existing measurement data}

Especially quantum algorithms using randomized classical shadows~\cite{huangPredictingManyProperties2020c,elbenRandomizedMeasurementToolbox2023, morrisSelectiveQuantumState2020,painiEstimatingExpectationValues2021,Efficient} are of particular interest. Since these strategies by default aim to estimate a set of $m$ $n$-qubit Pauli operators $P_i$, they constitute the prototypical ``measure now, compute functions of interest later" mindset. 
Thus, even if the main objective of a quantum algorithm differs from those discussed here, one still has the measurement data to reuse and can formulate a sampling error bound given the number of samples.

The performances of these methods are shown mostly heuristically with an analytical upper bound on the number of measurements given by Ref.~\cite[Theorem 3]{evansScalableBayesianHamiltonian2019}. 
It essentially states that for an error $\epsilon\in(0,1)$ and failure probability $\delta$, it suffices to (sequentially) use 
\begin{equation}
N=\frac{2}{\epsilon^2(1-\epsilon)}3^{w_\mathrm{max}}\log\frac{3m_\mathrm{tot}}{\delta}
\end{equation}
copies of $\rho$ to approximate each $\trace{P_i\rho}$ up to (additive) error $\epsilon$, where $w_\mathrm{max}$ denotes the maximum Pauli weight and $m_\mathrm{tot}$ denotes the number of Pauli expectation values to be estimated. 

For the presented algorithm, these results lead to the following conclusion: controlling the error of each $Q_i$ in Eq.~\eqref{eq:measurement_expansion} to $\epsilon/\norm{\gamma_i}_1$ and thus bounding the overall sampling error of $\widetilde{O}(t)$ by $\epsilon$ requires
\begin{equation}
    N=\mathcal{O}\left(3^{w_\mathrm{max}}\frac{\norm{\gamma}_1^2\log(m_\mathrm{tot}/\delta)}{\epsilon^2}\right)
\end{equation}
copies of $\rho$.
From Eq.~\eqref{eq:nr_of_terms}, we know that the number of expectation values to be estimated $m_\mathrm{tot}$ scales in the worst case (slightly) doubly exponentially with the system size, meaning that the system size is a limiting factor not only due to its influence on $\norm{\gamma}_1$.

While the actual performance could be better by orders of magnitude, these methods usually perform best for many low-weight or few global terms in contrast to the Taylor series, which outputs many terms with intermediate to large Pauli weight.
Having obtained measurement data for another primary task, we can reuse the measurement data to secondarily estimate matrix functions $(H,t)\mapsto F(H,t)$ up to an explicitly determinable error.

\section{Concrete example and reducing terms}
\label{sec:heisenberg_example}
To gain better intuition of how the expansion of time dynamics into a linear combination of Pauli operators of Eq.~(\ref{eq:o_expansion})  might look for a specific Hamiltonian, let us consider a Heisenberg Hamiltonian with equally weighted interactions, i.e.,
\begin{equation}
    H=\sum_{i=1}^{n-1}J\left(X_iX_{i+1}+Y_iY_{i+1}+Z_iZ_{i+1}\right),
    \label{eq:heisenberg}
\end{equation}
which has three Pauli operators per site and thus a total number of $L=3(n-1)$ terms.

Naively, we have upper-bounded the number of terms present in the expansion of the $K$-th order truncated Taylor series of the time evolution operator as $m_\mathrm{tot} = \sum_{k=0}^K L^k = (L^{K+1}-1)/(L-1)$.
Also, since $\lambda=3nJ$, we have $\norm{\gamma}_1=\mathrm{poly}(\ee^{(n-1)Jt})$, explicitly showing that the system size has a dependence on the number of required samples and practically feasible simulation time.
However, since the Pauli operators square to the identity and form a group, many terms will collapse to the identity matrix or reproduce existing terms as early as in the second order.
Explicitly, for
\begin{align*}
H^2 =& \left(\sum_{i=1}^{n-1} \underset{h_i}{\underbrace{J\left(X_iX_{i+1}+Y_iY_{i+1}+Z_iZ_{i+1}\right)}}\right)^2 \\
=& \sum_{i,j=1}^{n-1}h_i h_j = \sum_{i=1}^{n-1}h_i^2 + \sum_{i<j}(h_ih_j+h_jh_i) 
\end{align*}
 we find that 
\begin{equation}
    \left(h_i\right)^2=3J^2\id-2J^2h_i,
\end{equation}
which, instead of adding nine new Pauli operators, does not contribute to the growth of operators to be measured at all. 
Indeed, the identity term is already present in the zeroth order and the second term in the first order, leading to an overall reduction by $9(n-1)$ terms.
Furthermore, terms of the form $X_iX_{i+1}X_jX_{j+1}$ for $i\neq j$ will appear twice in the second sum, further reducing the number of terms by $3n(n-1)$.

Thus, using algebraic properties of the Pauli operators ensures that the total number of contributing terms for a second-order Taylor series is smaller than the worst-case bound of $(27(n-1)^3-1)/(3(n-1)-1)$ by $(n+9)(n-1)$, leading to $m_\mathrm{tot}=8n^2-23n+16$ terms.
Note, however, that although these terms do not contribute to the number of Pauli operators, they still contribute their pre-factors.
These pre-factors $\gamma_j$ in Eq.~\eqref{eq:o_expansion} will then be sums of powers of $Jt$ and must be considered in the error analysis, ultimately contributing to the sample complexity, where contributions to the identity term and negative contributions reduce $\norm{\gamma}_1$ and thus allow for larger simulation times.

However, this constitutes only a modest asymptotic reduction that cannot be explicitly generalized.

Further reductions in the number of contributing operators might be achieved by using a group different from the Pauli group as the basis for expanding the Hamiltonian and observable.
Turning towards the multi-qubit Clifford group, we can simplify the Heisenberg Hamiltonian to
\begin{equation}
    H=\sum_{i=1}^{n-1}\frac{J}{2} \mathrm{SWAP}_{i,i+1} - (n-1)\frac{J}{2}\id,
\end{equation}
using $\mathrm{SWAP}_{i,j}=(\id+X_iX_j+Y_iY_j+Z_iZ_j)/2$, thereby reducing the number of terms in the Hamiltonian by a factor of three.
Finding appropriate approximations is thus essential in improving the practicality of this method.

In general, finding such decompositions is a hard task and thus not necessarily feasible, albeit a trick to be aware of when reducing circuit evaluations.
Additionally, non-Hermitian operators, such as the Clifford operators, can not be directly measured and require additional circuitry, such as the Hadamard test, to perform measurements.
As such, using different decompositions might be even more useful in the purely classical simulation regime.

\section{Discussion of potential applications}
The proposed method offers a tradeoff between a genuine quantum operation, such as time evolution, and a more involved classical post-processing stage with additional sample complexity.
This can be viewed as a hybrid quantum-classical algorithm that builds on the power of quantum measurements. 
Furthermore, any quantum computation will require measurements to output its result in a classically understandable language.
Thus, we will always have access to at least a few measurements that could enhance most quantum computations in classical post-processing. Alternatively, when implemented as a standalone algorithm, we can crank up the number of measurements to improve its error.
However, we think that this method is best applied in combination with a capable but limited quantum device.
There, it could be integrated by exhausting the capabilities of the device and enhancing the computational power by additional measurements and the application of the Taylor expansion of a suitable spectral function in post-processing.

\subsection{Short-time dynamics by measurement}

The first potential application lies in quantum simulation.
As a guiding example, we have already studied the use for estimating time dynamics of observables via measurements, reminiscent of classical cluster expansion techniques applied to arbitrary quantum states~\cite{wildClassicalSimulationShortTime2023}.
As such, any application of Cluster expansion, such as the Loschmidt echo $\trace{\rho\ee^{-\ii H t}}$, is feasible.
Overcoming the restriction to product states and moving toward a hybrid quantum-classical algorithm opens the door to new use cases.
For near-term quantum devices, where time evolution algorithms are limited due to device noise but additional measurements might be realistic, we can imagine combining time evolution on the quantum device with additional short-time evolution in classical post-processing to enhance the quantum device capabilities.

To be more concrete, we can imagine a quantum circuit preparing the state $\rho(t_1)=\ee^{-\ii H t_1}\rho_0\ee^{\ii H t_1}$ for the N\'eel state $\rho_0=\ketbra{0,1,0,1,\cdots}{0,1,0,1,\cdots}$ as initial quantum state and the Heisenberg Hamiltonian from Eq.~\eqref{eq:heisenberg}. 

We can then estimate the expectation value of the particle imbalance $O=\sum_i (-1)^i Z_i$ not only for $\rho(t_1)$ but also classically construct an approximation for $\trace{O\rho(t_1+t_2)}$ for $t_2>0$ using the presented method.

As such, we envision a Schrödinger evolution to be implemented on actual quantum hardware and a short-time Heisenberg evolution simulated classically via additional measurements.

Again, we stress that short times really mean short times since the sampling overhead grows as $\mathrm{poly}(\ee^{\lambda t})$.

As an intermediate regime between this hybrid algorithm and classical \emph{cluster expansion methods}, we can also envision an extended cluster expansion that is not restricted to product states but to the superposition of $r$ product states, thereby increasing the computational cost, but at the same time remaining in the purely classical regime.

\subsection{Short-time Hamiltonian learning and device verification}
Another example of practical short-time quantum simulating in post-processing is the task of device verification or Hamiltonian learning \cite{baireyLearningLocalHamiltonian2019,kokailSelfverifyingVariationalQuantum2019,carrascoTheoreticalExperimentalPerspectives2021,HamiltonianLearning}.
Given a quantum device with an underlying, unknown Hamiltonian, we envision a protocol in which a well-known initial state is prepared, and the system then evolves under its Hamiltonian for a short time before the state is measured. In classical post-processing, we can then evolve the system in negative time under a model Hamiltonian and check the statistics on certain observables to test the error of the model Hamiltonian.

More concretely, we envision a protocol motivated by quantum \emph{process tomography} and \emph{randomized benchmarking}~\cite{klieschGuaranteedRecoveryQuantum2019,eisertQuantumCertificationBenchmarking2020a,klieschTheoryQuantumSystem2021a} in which we first prepare a basis state, e.g., $\rho=\ketbra{0}{0}$.
Then, we let the system naturally evolve the state as 
\begin{equation}
\rho(t)=\ee^{-\ii H_\mathrm{sys} t}\rho \ee^{\ii H_\mathrm{sys} t}.
\end{equation}
We can then gauge the error of a guess $H_\mathrm{guess}$ of the system's Hamiltonian $H_\mathrm{sys}$ with the presented method using $\trace{Z(-t)\rho(t)}$ for $Z(-t)=\ee^{-\ii H_\mathrm{guess}} Z \ee^{-\ii H_\mathrm{guess}}$, which would return zero if $H_{\mathrm{sys}}=H_{\mathrm{guess}}$.
Repeating this procedure for different input states and observables then provides information about the error of $H_\mathrm{guess}$ 

As such, it could be a quick device verification protocol restricted due to the short simulation times and less demanding on the hardware than extensive benchmarking protocols since it only relies on measurements and does not require additional quantum gates.

This method is, therefore, in the same spirit as Hamiltonian learning and device verification protocols using measurements of stationary states~\cite{baireyLearningLocalHamiltonian2019,kokailSelfverifyingVariationalQuantum2019,carrascoTheoreticalExperimentalPerspectives2021}. 
These use observables measured on states and their time evolution to construct an equation system from which the Pauli coefficients of the system's Hamiltonian can be reconstructed.
While these methods are tailored towards Hamiltonian learning, the proposed application of our Taylor series approach lies more in quick device verification.

\subsection{Short-time imaginary time evolution}
Apart from real-time evolution, the nested commutator relations also hold for imaginary time evolution, as does the requirement on the truncation order (see, e.g.,  Ref.~\cite[Lemma 3.2]{brandaoFasterQuantumClassical2022}).
Even though imaginary time evolution is mostly interesting for long simulation times due to its ability to drive a state toward its ground state, even a short simulation can be useful. In a setting where the initial state is prepared in a black-box fashion, e.g., a variational quantum algorithm or the output of some applied shortcut to adiabaticity, even short-time imaginary time evolution will push the prepared state closer toward the system's ground state.
As such, we can, with additional measurements, obtain measurement statistics that are closer to that of the true ground state in classical post-processing.

In more concrete terms, we can envision a trial state $\rho_\mathrm{trial}$ whose energy $\trace{H\rho_\mathrm{trial}}$ is of interest and supposed to approximate the ground state energy $\trace{H\rho_\mathrm{gs}}$.
Then, we can use additional measurements to estimate 
$(1/Z)\trace{H\ee^{-\tau H}\rho_\mathrm{trial}\ee^{-\tau H}}$ with $Z=\trace{\ee^{-2\tau H}}$, which is more representative of the ground state energy.

Further, since the 1-norm of the pre-factors of the expansion $\norm{\gamma}_1$ is unaffected by moving from real to imaginary time, the sampling overheads are identical for both cases.

\subsection{Enhancing sub-universal quantum simulators}
The underlying principle of this method is that both the generators of the time evolution operator and the observable belong to the same group.
The limitation to short simulation times further means that we are restricted to simulating operators close to the identity.
However, in contexts where one cannot perform arbitrary quantum operations, such as sub-universal quantum simulators, even operations close to the identity might be of interest if they cannot be implemented natively.
Therefore, as long as we can measure the generators of a unitary close to the identity and those generators belong to the same group as the observable of interest, we can apply this unitary in classical post-processing and thus expand the capabilities of the quantum simulator.

For example, imagine a quantum device that can only perform gates on neighboring qubits but measure arbitrary single-qubit Pauli operators.
Then, we can add long-range interactions in classical post-processing by, e.g., employing the presented method for $F(t) = \ee^{\ii H_1 t_1}\cdots \ee^{\ii H_r t_r}O \ee^{-\ii H_r t_r}\cdots \ee^{-\ii H_1 t_1}$, where the $H_i$ represent shallow sums of geometrically non-local Hermitian operators.

Together with the combination of imaginary time evolution with shortcuts to adiabaticity or device verification, these applications motivate classical post-processing techniques for analog or sub-universal quantum simulators.
For such applications, we stress that focusing on Pauli operators is unnecessary and that, e.g., fermionic Hamiltonians and fermionic observables can be combined similarly.

\section{Conclusion and outlook}

The introduced Taylor series method to estimate short-time dynamics via measurements constitutes another feasible reading of the ``measure now, compute functions of interest later'' mindset, which combines quantum computations with classical post-processing of measurement statistics. 
Classical shadows once more present themselves as a multi-faceted tool that can help estimate several unrelated properties simultaneously, enabling the search for new post-processing techniques, in this case allowing for a short-time quantum simulation through measurement application.
We view this method as an enhancement of quantum processors limited in their computational power, e.g., 
by combining a classically difficult computation such as time evolution where the maximal evolution time is limited by the quantum device with an additional short-time evolution through additional measurements.

Although limited to short simulation times, it provides further arguments for the power of \emph{quantum measurements}, adding to recent insights on the power of Pauli shadows on initial, untrained states in the context of quantum neural networks~\cite{bermejoQuantumConvolutionalNeural2024}. 
This is in line with other recent developments that aim to flesh out the power of quantum measurement in quantum state engineering, quantum computation, and quantum simulation. 
It motivates the further development of \emph{hybrid quantum-classical algorithms} and new ways to enhance them in post-processing.
The limitation to short times is not unexpected since it is present in the purely classical regime of cluster expansion techniques and due to the classical hardness of time evolution in general.

The focus on Pauli operators is not required, and decompositions of Hamiltonians into other operators following a group structure, such as Clifford operators, provides another path toward reducing resource costs and simplifying post-processing tasks. 
We hope the present work stimulates further rigorous research on quantum dynamics and hybrid quantum-classical algorithms.

\paragraph*{Note added:}
During the review process of this manuscript, we learned of a complementary framework by Fuller \emph{et al.}~\cite{fullerImprovedQuantumComputation2025}, which appeared after our first preprint and discusses the integration of classical simulation of backpropagated Heisenberg evolution with Schrödinger evolution executed on quantum hardware. 
While a detailed comparison of both approaches is of great practical interest, it is outside the scope of this manuscript.

\paragraph*{Acknowledgments:}
We thank Mark Steudtner, Kianna Wan, Earl Campbell, Mario Berta, Jose Carrasco, and Alvaro M. Alhambra for useful discussions. The Berlin team has 
been supported by the BMWK (PlanQK), the DFG (CRC 183), the BMBF (QPIC-1, MuniQC-Atoms, HYBRID), the Quantum Flagship (Millenion, Pasquans2),  
QuantERA (HQCC) and the ERC (DebuQC). RiK has been supported by the Austrian Science fund via SFB BeyondC (Grant-DOI 10.55776/F71), the projects QuantumReady (FFG 896217) and High-Performance integrated Quantum Computing (FFG 897481) within Quantum Austria (both managed by the FFG), as well as the ERC (q-shadows).

\bibliography{main.bib}

\begin{thebibliography}{62}%
\makeatletter
\providecommand \@ifxundefined [1]{%
 \@ifx{#1\undefined}
}%
\providecommand \@ifnum [1]{%
 \ifnum #1\expandafter \@firstoftwo
 \else \expandafter \@secondoftwo
 \fi
}%
\providecommand \@ifx [1]{%
 \ifx #1\expandafter \@firstoftwo
 \else \expandafter \@secondoftwo
 \fi
}%
\providecommand \natexlab [1]{#1}%
\providecommand \enquote  [1]{``#1''}%
\providecommand \bibnamefont  [1]{#1}%
\providecommand \bibfnamefont [1]{#1}%
\providecommand \citenamefont [1]{#1}%
\providecommand \href@noop [0]{\@secondoftwo}%
\providecommand \href [0]{\begingroup \@sanitize@url \@href}%
\providecommand \@href[1]{\@@startlink{#1}\@@href}%
\providecommand \@@href[1]{\endgroup#1\@@endlink}%
\providecommand \@sanitize@url [0]{\catcode `\\12\catcode `\$12\catcode `\&12\catcode `\#12\catcode `\^12\catcode `\_12\catcode `\%12\relax}%
\providecommand \@@startlink[1]{}%
\providecommand \@@endlink[0]{}%
\providecommand \url  [0]{\begingroup\@sanitize@url \@url }%
\providecommand \@url [1]{\endgroup\@href {#1}{\urlprefix }}%
\providecommand \urlprefix  [0]{URL }%
\providecommand \Eprint [0]{\href }%
\providecommand \doibase [0]{https://doi.org/}%
\providecommand \selectlanguage [0]{\@gobble}%
\providecommand \bibinfo  [0]{\@secondoftwo}%
\providecommand \bibfield  [0]{\@secondoftwo}%
\providecommand \translation [1]{[#1]}%
\providecommand \BibitemOpen [0]{}%
\providecommand \bibitemStop [0]{}%
\providecommand \bibitemNoStop [0]{.\EOS\space}%
\providecommand \EOS [0]{\spacefactor3000\relax}%
\providecommand \BibitemShut  [1]{\csname bibitem#1\endcsname}%
\let\auto@bib@innerbib\@empty
%</preamble>
\bibitem [{\citenamefont {Kim}\ \emph {et~al.}(2023)\citenamefont {Kim}, \citenamefont {Eddins}, \citenamefont {Anand}, \citenamefont {Wei}, \citenamefont {{van den Berg}}, \citenamefont {Rosenblatt}, \citenamefont {Nayfeh}, \citenamefont {Wu}, \citenamefont {Zaletel}, \citenamefont {Temme},\ and\ \citenamefont {Kandala}}]{kimEvidenceUtilityQuantum2023}%
  \BibitemOpen
  \bibfield  {author} {\bibinfo {author} {\bibfnamefont {Y.}~\bibnamefont {Kim}}, \bibinfo {author} {\bibfnamefont {A.}~\bibnamefont {Eddins}}, \bibinfo {author} {\bibfnamefont {S.}~\bibnamefont {Anand}}, \bibinfo {author} {\bibfnamefont {K.~X.}\ \bibnamefont {Wei}}, \bibinfo {author} {\bibfnamefont {E.}~\bibnamefont {{van den Berg}}}, \bibinfo {author} {\bibfnamefont {S.}~\bibnamefont {Rosenblatt}}, \bibinfo {author} {\bibfnamefont {H.}~\bibnamefont {Nayfeh}}, \bibinfo {author} {\bibfnamefont {Y.}~\bibnamefont {Wu}}, \bibinfo {author} {\bibfnamefont {M.}~\bibnamefont {Zaletel}}, \bibinfo {author} {\bibfnamefont {K.}~\bibnamefont {Temme}},\ and\ \bibinfo {author} {\bibfnamefont {A.}~\bibnamefont {Kandala}},\ }\bibfield  {title} {\bibinfo {title} {Evidence for the utility of quantum computing before fault tolerance},\ }\href {https://doi.org/10.1038/s41586-023-06096-3} {\bibfield  {journal} {\bibinfo  {journal} {Nature}\ }\textbf {\bibinfo {volume} {618}},\ \bibinfo {pages} {500} (\bibinfo {year}
  {2023})}\BibitemShut {NoStop}%
\bibitem [{\citenamefont {Reichardt}\ \emph {et~al.}(2024{\natexlab{a}})\citenamefont {Reichardt}, \citenamefont {Aasen}, \citenamefont {Chao}, \citenamefont {Chernoguzov}, \citenamefont {van Dam}, \citenamefont {Gaebler}, \citenamefont {Gresh}, \citenamefont {Lucchetti}, \citenamefont {Mills}, \citenamefont {Moses}, \citenamefont {Neyenhuis}, \citenamefont {Paetznick}, \citenamefont {Paz}, \citenamefont {Siegfried}, \citenamefont {da~Silva}, \citenamefont {Svore}, \citenamefont {Wang},\ and\ \citenamefont {Zanner}}]{reichardtDemonstrationQuantumComputation2024}%
  \BibitemOpen
  \bibfield  {author} {\bibinfo {author} {\bibfnamefont {B.~W.}\ \bibnamefont {Reichardt}}, \bibinfo {author} {\bibfnamefont {D.}~\bibnamefont {Aasen}}, \bibinfo {author} {\bibfnamefont {R.}~\bibnamefont {Chao}}, \bibinfo {author} {\bibfnamefont {A.}~\bibnamefont {Chernoguzov}}, \bibinfo {author} {\bibfnamefont {W.}~\bibnamefont {van Dam}}, \bibinfo {author} {\bibfnamefont {J.~P.}\ \bibnamefont {Gaebler}}, \bibinfo {author} {\bibfnamefont {D.}~\bibnamefont {Gresh}}, \bibinfo {author} {\bibfnamefont {D.}~\bibnamefont {Lucchetti}}, \bibinfo {author} {\bibfnamefont {M.}~\bibnamefont {Mills}}, \bibinfo {author} {\bibfnamefont {S.~A.}\ \bibnamefont {Moses}}, \bibinfo {author} {\bibfnamefont {B.}~\bibnamefont {Neyenhuis}}, \bibinfo {author} {\bibfnamefont {A.}~\bibnamefont {Paetznick}}, \bibinfo {author} {\bibfnamefont {A.}~\bibnamefont {Paz}}, \bibinfo {author} {\bibfnamefont {P.~E.}\ \bibnamefont {Siegfried}}, \bibinfo {author} {\bibfnamefont {M.~P.}\ \bibnamefont {da~Silva}}, \bibinfo {author} {\bibfnamefont
  {K.~M.}\ \bibnamefont {Svore}}, \bibinfo {author} {\bibfnamefont {Z.}~\bibnamefont {Wang}},\ and\ \bibinfo {author} {\bibfnamefont {M.}~\bibnamefont {Zanner}},\ }\bibfield  {title} {\bibinfo {title} {Demonstration of quantum computation and error correction with a tesseract code},\ }\href {https://doi.org/10.48550/arXiv.2409.04628} {\bibfield  {journal} {\bibinfo  {journal} {arXiv:2409.04628}\ } (\bibinfo {year} {2024}{\natexlab{a}})}\BibitemShut {NoStop}%
\bibitem [{\citenamefont {{Google Quantum AI and Collaborators}}(2024)}]{acharyaQuantumErrorCorrection2024}%
  \BibitemOpen
  \bibfield  {author} {\bibinfo {author} {\bibnamefont {{Google Quantum AI and Collaborators}}},\ }\bibfield  {title} {\bibinfo {title} {Quantum error correction below the surface code threshold},\ }\href {https://doi.org/10.1038/s41586-024-08449-y} {\bibfield  {journal} {\bibinfo  {journal} {Nature}\ ,\ \bibinfo {pages} {1}} (\bibinfo {year} {2024})}\BibitemShut {NoStop}%
\bibitem [{\citenamefont {Reichardt}\ \emph {et~al.}(2024{\natexlab{b}})\citenamefont {Reichardt} \emph {et~al.}}]{reichardtLogicalComputationDemonstrated2024}%
  \BibitemOpen
  \bibfield  {author} {\bibinfo {author} {\bibfnamefont {B.~W.}\ \bibnamefont {Reichardt}} \emph {et~al.},\ }\bibfield  {title} {\bibinfo {title} {Logical computation demonstrated with a neutral atom quantum processor},\ }\href {https://doi.org/10.48550/arXiv.2411.11822} {\bibfield  {journal} {\bibinfo  {journal} {arXiv:2411.11822}\ } (\bibinfo {year} {2024}{\natexlab{b}})}\BibitemShut {NoStop}%
\bibitem [{\citenamefont {Bluvstein}\ \emph {et~al.}(2024)\citenamefont {Bluvstein}, \citenamefont {Evered}, \citenamefont {Geim}, \citenamefont {Li}, \citenamefont {Zhou}, \citenamefont {Manovitz}, \citenamefont {Ebadi}, \citenamefont {Cain}, \citenamefont {Kalinowski}, \citenamefont {Hangleiter}, \citenamefont {Bonilla~Ataides}, \citenamefont {Maskara}, \citenamefont {Cong}, \citenamefont {Gao}, \citenamefont {Sales~Rodriguez}, \citenamefont {Karolyshyn}, \citenamefont {Semeghini}, \citenamefont {Gullans}, \citenamefont {Greiner}, \citenamefont {Vuleti{\'c}},\ and\ \citenamefont {Lukin}}]{bluvsteinLogicalQuantumProcessor2024}%
  \BibitemOpen
  \bibfield  {author} {\bibinfo {author} {\bibfnamefont {D.}~\bibnamefont {Bluvstein}}, \bibinfo {author} {\bibfnamefont {S.~J.}\ \bibnamefont {Evered}}, \bibinfo {author} {\bibfnamefont {A.~A.}\ \bibnamefont {Geim}}, \bibinfo {author} {\bibfnamefont {S.~H.}\ \bibnamefont {Li}}, \bibinfo {author} {\bibfnamefont {H.}~\bibnamefont {Zhou}}, \bibinfo {author} {\bibfnamefont {T.}~\bibnamefont {Manovitz}}, \bibinfo {author} {\bibfnamefont {S.}~\bibnamefont {Ebadi}}, \bibinfo {author} {\bibfnamefont {M.}~\bibnamefont {Cain}}, \bibinfo {author} {\bibfnamefont {M.}~\bibnamefont {Kalinowski}}, \bibinfo {author} {\bibfnamefont {D.}~\bibnamefont {Hangleiter}}, \bibinfo {author} {\bibfnamefont {J.~P.}\ \bibnamefont {Bonilla~Ataides}}, \bibinfo {author} {\bibfnamefont {N.}~\bibnamefont {Maskara}}, \bibinfo {author} {\bibfnamefont {I.}~\bibnamefont {Cong}}, \bibinfo {author} {\bibfnamefont {X.}~\bibnamefont {Gao}}, \bibinfo {author} {\bibfnamefont {P.}~\bibnamefont {Sales~Rodriguez}}, \bibinfo {author} {\bibfnamefont
  {T.}~\bibnamefont {Karolyshyn}}, \bibinfo {author} {\bibfnamefont {G.}~\bibnamefont {Semeghini}}, \bibinfo {author} {\bibfnamefont {M.~J.}\ \bibnamefont {Gullans}}, \bibinfo {author} {\bibfnamefont {M.}~\bibnamefont {Greiner}}, \bibinfo {author} {\bibfnamefont {V.}~\bibnamefont {Vuleti{\'c}}},\ and\ \bibinfo {author} {\bibfnamefont {M.~D.}\ \bibnamefont {Lukin}},\ }\bibfield  {title} {\bibinfo {title} {Logical quantum processor based on reconfigurable atom arrays},\ }\href {https://doi.org/10.1038/s41586-023-06927-3} {\bibfield  {journal} {\bibinfo  {journal} {Nature}\ }\textbf {\bibinfo {volume} {626}},\ \bibinfo {pages} {58} (\bibinfo {year} {2024})}\BibitemShut {NoStop}%
\bibitem [{\citenamefont {Joshi}\ \emph {et~al.}(2023)\citenamefont {Joshi}, \citenamefont {Kokail}, \citenamefont {{van Bijnen}}, \citenamefont {Kranzl}, \citenamefont {Zache}, \citenamefont {Blatt}, \citenamefont {Roos},\ and\ \citenamefont {Zoller}}]{joshiExploringLargescaleEntanglement2023}%
  \BibitemOpen
  \bibfield  {author} {\bibinfo {author} {\bibfnamefont {M.~K.}\ \bibnamefont {Joshi}}, \bibinfo {author} {\bibfnamefont {C.}~\bibnamefont {Kokail}}, \bibinfo {author} {\bibfnamefont {R.}~\bibnamefont {{van Bijnen}}}, \bibinfo {author} {\bibfnamefont {F.}~\bibnamefont {Kranzl}}, \bibinfo {author} {\bibfnamefont {T.~V.}\ \bibnamefont {Zache}}, \bibinfo {author} {\bibfnamefont {R.}~\bibnamefont {Blatt}}, \bibinfo {author} {\bibfnamefont {C.~F.}\ \bibnamefont {Roos}},\ and\ \bibinfo {author} {\bibfnamefont {P.}~\bibnamefont {Zoller}},\ }\bibfield  {title} {\bibinfo {title} {Exploring large-scale entanglement in quantum simulation},\ }\href {https://doi.org/10.1038/s41586-023-06768-0} {\bibfield  {journal} {\bibinfo  {journal} {Nature}\ }\textbf {\bibinfo {volume} {624}},\ \bibinfo {pages} {539} (\bibinfo {year} {2023})}\BibitemShut {NoStop}%
\bibitem [{\citenamefont {Vollbrecht}\ and\ \citenamefont {Cirac}(2008)}]{Vollbrecht}%
  \BibitemOpen
  \bibfield  {author} {\bibinfo {author} {\bibfnamefont {K.~G.~H.}\ \bibnamefont {Vollbrecht}}\ and\ \bibinfo {author} {\bibfnamefont {J.~I.}\ \bibnamefont {Cirac}},\ }\bibfield  {title} {\bibinfo {title} {Quantum simulators, continuous-time automata, and translationally invariant systems},\ }\href {https://doi.org/10.1103/PhysRevLett.100.010501} {\bibfield  {journal} {\bibinfo  {journal} {Phys. Rev. Lett.}\ }\textbf {\bibinfo {volume} {100}},\ \bibinfo {pages} {010501} (\bibinfo {year} {2008})}\BibitemShut {NoStop}%
\bibitem [{\citenamefont {Baez}\ \emph {et~al.}(2020)\citenamefont {Baez}, \citenamefont {Goihl}, \citenamefont {Haferkamp}, \citenamefont {Bermejo-Vega}, \citenamefont {Gluza},\ and\ \citenamefont {Eisert}}]{DynamicalStructureFactors}%
  \BibitemOpen
  \bibfield  {author} {\bibinfo {author} {\bibfnamefont {M.~L.}\ \bibnamefont {Baez}}, \bibinfo {author} {\bibfnamefont {M.}~\bibnamefont {Goihl}}, \bibinfo {author} {\bibfnamefont {J.}~\bibnamefont {Haferkamp}}, \bibinfo {author} {\bibfnamefont {J.}~\bibnamefont {Bermejo-Vega}}, \bibinfo {author} {\bibfnamefont {M.}~\bibnamefont {Gluza}},\ and\ \bibinfo {author} {\bibfnamefont {J.}~\bibnamefont {Eisert}},\ }\bibfield  {title} {\bibinfo {title} {Dynamical structure factors of dynamical quantum simulators},\ }\href {https://doi.org/10.1073/pnas.2006103117} {\bibfield  {journal} {\bibinfo  {journal} {PNAS}\ }\textbf {\bibinfo {volume} {117}},\ \bibinfo {pages} {26123} (\bibinfo {year} {2020})}\BibitemShut {NoStop}%
\bibitem [{\citenamefont {Haah}\ \emph {et~al.}(2023)\citenamefont {Haah}, \citenamefont {Hastings}, \citenamefont {Kothari},\ and\ \citenamefont {Low}}]{BQPTimeEvolution}%
  \BibitemOpen
  \bibfield  {author} {\bibinfo {author} {\bibfnamefont {J.}~\bibnamefont {Haah}}, \bibinfo {author} {\bibfnamefont {M.~B.}\ \bibnamefont {Hastings}}, \bibinfo {author} {\bibfnamefont {R.}~\bibnamefont {Kothari}},\ and\ \bibinfo {author} {\bibfnamefont {G.~H.}\ \bibnamefont {Low}},\ }\bibfield  {title} {\bibinfo {title} {Quantum algorithm for simulating real time evolution of lattice hamiltonians},\ }\href {https://doi.org/10.1137/18M1231511} {\bibfield  {journal} {\bibinfo  {journal} {SIAM Journal on Computing}\ }\textbf {\bibinfo {volume} {52}},\ \bibinfo {pages} {FOCS18} (\bibinfo {year} {2023})}\BibitemShut {NoStop}%
\bibitem [{\citenamefont {Huang}\ \emph {et~al.}(2020)\citenamefont {Huang}, \citenamefont {Kueng},\ and\ \citenamefont {Preskill}}]{huangPredictingManyProperties2020c}%
  \BibitemOpen
  \bibfield  {author} {\bibinfo {author} {\bibfnamefont {H.-Y.}\ \bibnamefont {Huang}}, \bibinfo {author} {\bibfnamefont {R.}~\bibnamefont {Kueng}},\ and\ \bibinfo {author} {\bibfnamefont {J.}~\bibnamefont {Preskill}},\ }\bibfield  {title} {\bibinfo {title} {Predicting many properties of a quantum system from very few measurements},\ }\href {https://doi.org/10.1038/s41567-020-0932-7} {\bibfield  {journal} {\bibinfo  {journal} {Nature Phys.}\ }\textbf {\bibinfo {volume} {16}},\ \bibinfo {pages} {1050} (\bibinfo {year} {2020})}\BibitemShut {NoStop}%
\bibitem [{\citenamefont {Elben}\ \emph {et~al.}(2023)\citenamefont {Elben}, \citenamefont {Flammia}, \citenamefont {Huang}, \citenamefont {Kueng}, \citenamefont {Preskill}, \citenamefont {Vermersch},\ and\ \citenamefont {Zoller}}]{elbenRandomizedMeasurementToolbox2023}%
  \BibitemOpen
  \bibfield  {author} {\bibinfo {author} {\bibfnamefont {A.}~\bibnamefont {Elben}}, \bibinfo {author} {\bibfnamefont {S.~T.}\ \bibnamefont {Flammia}}, \bibinfo {author} {\bibfnamefont {H.-Y.}\ \bibnamefont {Huang}}, \bibinfo {author} {\bibfnamefont {R.}~\bibnamefont {Kueng}}, \bibinfo {author} {\bibfnamefont {J.}~\bibnamefont {Preskill}}, \bibinfo {author} {\bibfnamefont {B.}~\bibnamefont {Vermersch}},\ and\ \bibinfo {author} {\bibfnamefont {P.}~\bibnamefont {Zoller}},\ }\bibfield  {title} {\bibinfo {title} {The randomized measurement toolbox},\ }\href {https://doi.org/10.1038/s42254-022-00535-2} {\bibfield  {journal} {\bibinfo  {journal} {Nature Rev. Phys.}\ }\textbf {\bibinfo {volume} {5}},\ \bibinfo {pages} {9} (\bibinfo {year} {2023})}\BibitemShut {NoStop}%
\bibitem [{\citenamefont {Morris}\ and\ \citenamefont {Daki{\'c}}(2020)}]{morrisSelectiveQuantumState2020}%
  \BibitemOpen
  \bibfield  {author} {\bibinfo {author} {\bibfnamefont {J.}~\bibnamefont {Morris}}\ and\ \bibinfo {author} {\bibfnamefont {B.}~\bibnamefont {Daki{\'c}}},\ }\bibfield  {title} {\bibinfo {title} {Selective {{Quantum State Tomography}}},\ }\href {https://doi.org/10.48550/arXiv.1909.05880} {\bibfield  {journal} {\bibinfo  {journal} {arXiv:1909.05880}\ } (\bibinfo {year} {2020})}\BibitemShut {NoStop}%
\bibitem [{\citenamefont {Paini}\ \emph {et~al.}(2021)\citenamefont {Paini}, \citenamefont {Kalev}, \citenamefont {Padilha},\ and\ \citenamefont {Ruck}}]{painiEstimatingExpectationValues2021}%
  \BibitemOpen
  \bibfield  {author} {\bibinfo {author} {\bibfnamefont {M.}~\bibnamefont {Paini}}, \bibinfo {author} {\bibfnamefont {A.}~\bibnamefont {Kalev}}, \bibinfo {author} {\bibfnamefont {D.}~\bibnamefont {Padilha}},\ and\ \bibinfo {author} {\bibfnamefont {B.}~\bibnamefont {Ruck}},\ }\bibfield  {title} {\bibinfo {title} {Estimating expectation values using approximate quantum states},\ }\href {https://doi.org/10.22331/q-2021-03-16-413} {\bibfield  {journal} {\bibinfo  {journal} {Quantum}\ }\textbf {\bibinfo {volume} {5}},\ \bibinfo {pages} {413} (\bibinfo {year} {2021})}\BibitemShut {NoStop}%
\bibitem [{\citenamefont {Peruzzo}\ \emph {et~al.}(2014)\citenamefont {Peruzzo}, \citenamefont {McClean}, \citenamefont {Shadbolt}, \citenamefont {Yung}, \citenamefont {Zhou}, \citenamefont {Love}, \citenamefont {{Aspuru-Guzik}},\ and\ \citenamefont {O'Brien}}]{peruzzoVariationalEigenvalueSolver2014}%
  \BibitemOpen
  \bibfield  {author} {\bibinfo {author} {\bibfnamefont {A.}~\bibnamefont {Peruzzo}}, \bibinfo {author} {\bibfnamefont {J.}~\bibnamefont {McClean}}, \bibinfo {author} {\bibfnamefont {P.}~\bibnamefont {Shadbolt}}, \bibinfo {author} {\bibfnamefont {M.-H.}\ \bibnamefont {Yung}}, \bibinfo {author} {\bibfnamefont {X.-Q.}\ \bibnamefont {Zhou}}, \bibinfo {author} {\bibfnamefont {P.~J.}\ \bibnamefont {Love}}, \bibinfo {author} {\bibfnamefont {A.}~\bibnamefont {{Aspuru-Guzik}}},\ and\ \bibinfo {author} {\bibfnamefont {J.~L.}\ \bibnamefont {O'Brien}},\ }\bibfield  {title} {\bibinfo {title} {A variational eigenvalue solver on a photonic quantum processor},\ }\href {https://doi.org/10.1038/ncomms5213} {\bibfield  {journal} {\bibinfo  {journal} {Nature Communications}\ }\textbf {\bibinfo {volume} {5}},\ \bibinfo {pages} {4213} (\bibinfo {year} {2014})}\BibitemShut {NoStop}%
\bibitem [{\citenamefont {Bharti}\ \emph {et~al.}(2022)\citenamefont {Bharti}, \citenamefont {{Cervera-Lierta}}, \citenamefont {Kyaw}, \citenamefont {Haug}, \citenamefont {{Alperin-Lea}}, \citenamefont {Anand}, \citenamefont {Degroote}, \citenamefont {Heimonen}, \citenamefont {Kottmann}, \citenamefont {Menke}, \citenamefont {Mok}, \citenamefont {Sim}, \citenamefont {Kwek},\ and\ \citenamefont {{Aspuru-Guzik}}}]{bhartiNoisyIntermediatescaleQuantum2022}%
  \BibitemOpen
  \bibfield  {author} {\bibinfo {author} {\bibfnamefont {K.}~\bibnamefont {Bharti}}, \bibinfo {author} {\bibfnamefont {A.}~\bibnamefont {{Cervera-Lierta}}}, \bibinfo {author} {\bibfnamefont {T.~H.}\ \bibnamefont {Kyaw}}, \bibinfo {author} {\bibfnamefont {T.}~\bibnamefont {Haug}}, \bibinfo {author} {\bibfnamefont {S.}~\bibnamefont {{Alperin-Lea}}}, \bibinfo {author} {\bibfnamefont {A.}~\bibnamefont {Anand}}, \bibinfo {author} {\bibfnamefont {M.}~\bibnamefont {Degroote}}, \bibinfo {author} {\bibfnamefont {H.}~\bibnamefont {Heimonen}}, \bibinfo {author} {\bibfnamefont {J.~S.}\ \bibnamefont {Kottmann}}, \bibinfo {author} {\bibfnamefont {T.}~\bibnamefont {Menke}}, \bibinfo {author} {\bibfnamefont {W.-K.}\ \bibnamefont {Mok}}, \bibinfo {author} {\bibfnamefont {S.}~\bibnamefont {Sim}}, \bibinfo {author} {\bibfnamefont {L.-C.}\ \bibnamefont {Kwek}},\ and\ \bibinfo {author} {\bibfnamefont {A.}~\bibnamefont {{Aspuru-Guzik}}},\ }\bibfield  {title} {\bibinfo {title} {Noisy intermediate-scale quantum algorithms},\ }\href
  {https://doi.org/10.1103/RevModPhys.94.015004} {\bibfield  {journal} {\bibinfo  {journal} {Rev. Mod. Phys.}\ }\textbf {\bibinfo {volume} {94}},\ \bibinfo {pages} {015004} (\bibinfo {year} {2022})}\BibitemShut {NoStop}%
\bibitem [{\citenamefont {Cerezo}\ \emph {et~al.}(2021)\citenamefont {Cerezo}, \citenamefont {Arrasmith}, \citenamefont {Babbush}, \citenamefont {Benjamin}, \citenamefont {Endo}, \citenamefont {Fujii}, \citenamefont {McClean}, \citenamefont {Mitarai}, \citenamefont {Yuan}, \citenamefont {Cincio},\ and\ \citenamefont {Coles}}]{cerezoVariationalQuantumAlgorithms2021}%
  \BibitemOpen
  \bibfield  {author} {\bibinfo {author} {\bibfnamefont {M.}~\bibnamefont {Cerezo}}, \bibinfo {author} {\bibfnamefont {A.}~\bibnamefont {Arrasmith}}, \bibinfo {author} {\bibfnamefont {R.}~\bibnamefont {Babbush}}, \bibinfo {author} {\bibfnamefont {S.~C.}\ \bibnamefont {Benjamin}}, \bibinfo {author} {\bibfnamefont {S.}~\bibnamefont {Endo}}, \bibinfo {author} {\bibfnamefont {K.}~\bibnamefont {Fujii}}, \bibinfo {author} {\bibfnamefont {J.~R.}\ \bibnamefont {McClean}}, \bibinfo {author} {\bibfnamefont {K.}~\bibnamefont {Mitarai}}, \bibinfo {author} {\bibfnamefont {X.}~\bibnamefont {Yuan}}, \bibinfo {author} {\bibfnamefont {L.}~\bibnamefont {Cincio}},\ and\ \bibinfo {author} {\bibfnamefont {P.~J.}\ \bibnamefont {Coles}},\ }\bibfield  {title} {\bibinfo {title} {Variational {{quantum algorithms}}},\ }\href {https://doi.org/10.1038/s42254-021-00348-9} {\bibfield  {journal} {\bibinfo  {journal} {Nature Rev. Phys.}\ }\textbf {\bibinfo {volume} {3}},\ \bibinfo {pages} {625} (\bibinfo {year} {2021})}\BibitemShut {NoStop}%
\bibitem [{\citenamefont {Bertoni}\ \emph {et~al.}(2024)\citenamefont {Bertoni}, \citenamefont {Haferkamp}, \citenamefont {Hinsche}, \citenamefont {Ioannou}, \citenamefont {Eisert},\ and\ \citenamefont {Pashayan}}]{bertoniShallowShadowsExpectation2024}%
  \BibitemOpen
  \bibfield  {author} {\bibinfo {author} {\bibfnamefont {C.}~\bibnamefont {Bertoni}}, \bibinfo {author} {\bibfnamefont {J.}~\bibnamefont {Haferkamp}}, \bibinfo {author} {\bibfnamefont {M.}~\bibnamefont {Hinsche}}, \bibinfo {author} {\bibfnamefont {M.}~\bibnamefont {Ioannou}}, \bibinfo {author} {\bibfnamefont {J.}~\bibnamefont {Eisert}},\ and\ \bibinfo {author} {\bibfnamefont {H.}~\bibnamefont {Pashayan}},\ }\bibfield  {title} {\bibinfo {title} {Shallow {{shadows}}: {{Expectation estimation using low-depth random Clifford circuits}}},\ }\href {https://doi.org/10.1103/PhysRevLett.133.020602} {\bibfield  {journal} {\bibinfo  {journal} {Phys. Rev. Lett.}\ }\textbf {\bibinfo {volume} {133}},\ \bibinfo {pages} {020602} (\bibinfo {year} {2024})}\BibitemShut {NoStop}%
\bibitem [{\citenamefont {Chan}\ \emph {et~al.}(2022)\citenamefont {Chan}, \citenamefont {Meister}, \citenamefont {Goh},\ and\ \citenamefont {Koczor}}]{chanAlgorithmicShadowSpectroscopy2024}%
  \BibitemOpen
  \bibfield  {author} {\bibinfo {author} {\bibfnamefont {H.~H.~S.}\ \bibnamefont {Chan}}, \bibinfo {author} {\bibfnamefont {R.}~\bibnamefont {Meister}}, \bibinfo {author} {\bibfnamefont {M.~L.}\ \bibnamefont {Goh}},\ and\ \bibinfo {author} {\bibfnamefont {B.}~\bibnamefont {Koczor}},\ }\bibfield  {title} {\bibinfo {title} {Algorithmic {{Shadow Spectroscopy}}},\ }\href {https://doi.org/10.48550/arXiv.2212.11036} {\bibfield  {journal} {\bibinfo  {journal} {arXiv:2212.11036}\ } (\bibinfo {year} {2022})}\BibitemShut {NoStop}%
\bibitem [{\citenamefont {Wu}\ \emph {et~al.}(2023)\citenamefont {Wu}, \citenamefont {Sun}, \citenamefont {Huang},\ and\ \citenamefont {Yuan}}]{wuOverlappedGroupingMeasurement2021}%
  \BibitemOpen
  \bibfield  {author} {\bibinfo {author} {\bibfnamefont {B.}~\bibnamefont {Wu}}, \bibinfo {author} {\bibfnamefont {J.}~\bibnamefont {Sun}}, \bibinfo {author} {\bibfnamefont {Q.}~\bibnamefont {Huang}},\ and\ \bibinfo {author} {\bibfnamefont {X.}~\bibnamefont {Yuan}},\ }\bibfield  {title} {\bibinfo {title} {Overlapped grouping measurement: {{A}} unified framework for measuring quantum states},\ }\href {https://doi.org/10.22331/q-2023-01-13-896} {\bibfield  {journal} {\bibinfo  {journal} {Quantum}\ }\textbf {\bibinfo {volume} {7}},\ \bibinfo {pages} {896} (\bibinfo {year} {2023})}\BibitemShut {NoStop}%
\bibitem [{\citenamefont {Verteletskyi}\ \emph {et~al.}(2020)\citenamefont {Verteletskyi}, \citenamefont {Yen},\ and\ \citenamefont {Izmaylov}}]{verteletskyiMeasurementOptimizationVariational2020}%
  \BibitemOpen
  \bibfield  {author} {\bibinfo {author} {\bibfnamefont {V.}~\bibnamefont {Verteletskyi}}, \bibinfo {author} {\bibfnamefont {T.-C.}\ \bibnamefont {Yen}},\ and\ \bibinfo {author} {\bibfnamefont {A.~F.}\ \bibnamefont {Izmaylov}},\ }\bibfield  {title} {\bibinfo {title} {Measurement {{optimization}} in the {{variational quantum eigensolver using}} a {{minimum clique cover}}},\ }\href {https://doi.org/10.1063/1.5141458} {\bibfield  {journal} {\bibinfo  {journal} {J. Chem. Phys.}\ }\textbf {\bibinfo {volume} {152}},\ \bibinfo {pages} {124114} (\bibinfo {year} {2020})}\BibitemShut {NoStop}%
\bibitem [{\citenamefont {Huang}\ \emph {et~al.}(2021)\citenamefont {Huang}, \citenamefont {Kueng},\ and\ \citenamefont {Preskill}}]{huangEfficientEstimationPauli2021}%
  \BibitemOpen
  \bibfield  {author} {\bibinfo {author} {\bibfnamefont {H.-Y.}\ \bibnamefont {Huang}}, \bibinfo {author} {\bibfnamefont {R.}~\bibnamefont {Kueng}},\ and\ \bibinfo {author} {\bibfnamefont {J.}~\bibnamefont {Preskill}},\ }\bibfield  {title} {\bibinfo {title} {Efficient {{estimation}} of {{Pauli observables}} by {{derandomization}}},\ }\href {https://doi.org/10.1103/PhysRevLett.127.030503} {\bibfield  {journal} {\bibinfo  {journal} {Phys. Rev. Lett.}\ }\textbf {\bibinfo {volume} {127}},\ \bibinfo {pages} {030503} (\bibinfo {year} {2021})}\BibitemShut {NoStop}%
\bibitem [{\citenamefont {Hadfield}\ \emph {et~al.}(2022)\citenamefont {Hadfield}, \citenamefont {Bravyi}, \citenamefont {Raymond},\ and\ \citenamefont {Mezzacapo}}]{hadfieldMeasurementsQuantumHamiltonians2022}%
  \BibitemOpen
  \bibfield  {author} {\bibinfo {author} {\bibfnamefont {C.}~\bibnamefont {Hadfield}}, \bibinfo {author} {\bibfnamefont {S.}~\bibnamefont {Bravyi}}, \bibinfo {author} {\bibfnamefont {R.}~\bibnamefont {Raymond}},\ and\ \bibinfo {author} {\bibfnamefont {A.}~\bibnamefont {Mezzacapo}},\ }\bibfield  {title} {\bibinfo {title} {Measurements of {{Quantum Hamiltonians}} with {{Locally-Biased Classical Shadows}}},\ }\href {https://doi.org/10.1007/s00220-022-04343-8} {\bibfield  {journal} {\bibinfo  {journal} {Communications in Mathematical Physics}\ }\textbf {\bibinfo {volume} {391}},\ \bibinfo {pages} {951} (\bibinfo {year} {2022})}\BibitemShut {NoStop}%
\bibitem [{\citenamefont {Campbell}(2017)}]{campbellShorterGateSequences2017b}%
  \BibitemOpen
  \bibfield  {author} {\bibinfo {author} {\bibfnamefont {E.}~\bibnamefont {Campbell}},\ }\bibfield  {title} {\bibinfo {title} {Shorter gate sequences for quantum computing by mixing unitaries},\ }\href {https://doi.org/10.1103/PhysRevA.95.042306} {\bibfield  {journal} {\bibinfo  {journal} {Phys. Rev. A}\ }\textbf {\bibinfo {volume} {95}},\ \bibinfo {pages} {042306} (\bibinfo {year} {2017})}\BibitemShut {NoStop}%
\bibitem [{\citenamefont {Campbell}(2019)}]{campbellRandomCompilerFast2019a}%
  \BibitemOpen
  \bibfield  {author} {\bibinfo {author} {\bibfnamefont {E.}~\bibnamefont {Campbell}},\ }\bibfield  {title} {\bibinfo {title} {A random compiler for fast {{Hamiltonian}} simulation},\ }\href {https://doi.org/10.1103/PhysRevLett.123.070503} {\bibfield  {journal} {\bibinfo  {journal} {Phys. Rev. Lett.}\ }\textbf {\bibinfo {volume} {123}},\ \bibinfo {pages} {070503} (\bibinfo {year} {2019})}\BibitemShut {NoStop}%
\bibitem [{\citenamefont {Childs}\ \emph {et~al.}(2019)\citenamefont {Childs}, \citenamefont {Ostrander},\ and\ \citenamefont {Su}}]{childsFasterQuantumSimulation2019}%
  \BibitemOpen
  \bibfield  {author} {\bibinfo {author} {\bibfnamefont {A.~M.}\ \bibnamefont {Childs}}, \bibinfo {author} {\bibfnamefont {A.}~\bibnamefont {Ostrander}},\ and\ \bibinfo {author} {\bibfnamefont {Y.}~\bibnamefont {Su}},\ }\bibfield  {title} {\bibinfo {title} {Faster quantum simulation by randomization},\ }\href {https://doi.org/10.22331/q-2019-09-02-182} {\bibfield  {journal} {\bibinfo  {journal} {Quantum}\ }\textbf {\bibinfo {volume} {3}},\ \bibinfo {pages} {182} (\bibinfo {year} {2019})}\BibitemShut {NoStop}%
\bibitem [{\citenamefont {Chen}\ \emph {et~al.}(2021)\citenamefont {Chen}, \citenamefont {Huang}, \citenamefont {Kueng},\ and\ \citenamefont {Tropp}}]{chenConcentrationRandomProduct2021}%
  \BibitemOpen
  \bibfield  {author} {\bibinfo {author} {\bibfnamefont {C.-F.}\ \bibnamefont {Chen}}, \bibinfo {author} {\bibfnamefont {H.-Y.}\ \bibnamefont {Huang}}, \bibinfo {author} {\bibfnamefont {R.}~\bibnamefont {Kueng}},\ and\ \bibinfo {author} {\bibfnamefont {J.~A.}\ \bibnamefont {Tropp}},\ }\bibfield  {title} {\bibinfo {title} {Concentration for {{random product formulas}}},\ }\href {https://doi.org/10.1103/PRXQuantum.2.040305} {\bibfield  {journal} {\bibinfo  {journal} {PRX Quantum}\ }\textbf {\bibinfo {volume} {2}},\ \bibinfo {pages} {040305} (\bibinfo {year} {2021})}\BibitemShut {NoStop}%
\bibitem [{\citenamefont {Faehrmann}\ \emph {et~al.}(2022)\citenamefont {Faehrmann}, \citenamefont {Steudtner}, \citenamefont {Kueng}, \citenamefont {Kieferova},\ and\ \citenamefont {Eisert}}]{faehrmannRandomizingMultiproductFormulas2022}%
  \BibitemOpen
  \bibfield  {author} {\bibinfo {author} {\bibfnamefont {P.~K.}\ \bibnamefont {Faehrmann}}, \bibinfo {author} {\bibfnamefont {M.}~\bibnamefont {Steudtner}}, \bibinfo {author} {\bibfnamefont {R.}~\bibnamefont {Kueng}}, \bibinfo {author} {\bibfnamefont {M.}~\bibnamefont {Kieferova}},\ and\ \bibinfo {author} {\bibfnamefont {J.}~\bibnamefont {Eisert}},\ }\bibfield  {title} {\bibinfo {title} {Randomizing multi-product formulas for {{Hamiltonian}} simulation},\ }\href {https://doi.org/10.22331/q-2022-09-19-806} {\bibfield  {journal} {\bibinfo  {journal} {Quantum}\ }\textbf {\bibinfo {volume} {6}},\ \bibinfo {pages} {806} (\bibinfo {year} {2022})}\BibitemShut {NoStop}%
\bibitem [{\citenamefont {Chakraborty}(2024)}]{chakrabortyImplementingAnyLinear2024}%
  \BibitemOpen
  \bibfield  {author} {\bibinfo {author} {\bibfnamefont {S.}~\bibnamefont {Chakraborty}},\ }\bibfield  {title} {\bibinfo {title} {Implementing any {{Linear Combination}} of {{Unitaries}} on {{Intermediate-term Quantum Computers}}},\ }\href {https://doi.org/10.22331/q-2024-10-10-1496} {\bibfield  {journal} {\bibinfo  {journal} {Quantum}\ }\textbf {\bibinfo {volume} {8}},\ \bibinfo {pages} {1496} (\bibinfo {year} {2024})}\BibitemShut {NoStop}%
\bibitem [{\citenamefont {Low}\ \emph {et~al.}(2019)\citenamefont {Low}, \citenamefont {Kliuchnikov},\ and\ \citenamefont {Wiebe}}]{lowWellconditionedMultiproductHamiltonian2019}%
  \BibitemOpen
  \bibfield  {author} {\bibinfo {author} {\bibfnamefont {G.~H.}\ \bibnamefont {Low}}, \bibinfo {author} {\bibfnamefont {V.}~\bibnamefont {Kliuchnikov}},\ and\ \bibinfo {author} {\bibfnamefont {N.}~\bibnamefont {Wiebe}},\ }\bibfield  {title} {\bibinfo {title} {Well-conditioned multiproduct {{Hamiltonian}} simulation},\ }\href {https://doi.org/10.48550/arXiv.1907.11679} {\bibfield  {journal} {\bibinfo  {journal} {arXiv:1907.11679}\ } (\bibinfo {year} {2019})}\BibitemShut {NoStop}%
\bibitem [{\citenamefont {Berry}\ \emph {et~al.}(2015{\natexlab{a}})\citenamefont {Berry}, \citenamefont {Childs},\ and\ \citenamefont {Kothari}}]{berryHamiltonianSimulationNearly2015}%
  \BibitemOpen
  \bibfield  {author} {\bibinfo {author} {\bibfnamefont {D.~W.}\ \bibnamefont {Berry}}, \bibinfo {author} {\bibfnamefont {A.~M.}\ \bibnamefont {Childs}},\ and\ \bibinfo {author} {\bibfnamefont {R.}~\bibnamefont {Kothari}},\ }\bibfield  {title} {\bibinfo {title} {Hamiltonian {{simulation}} with {{nearly optimal dependence}} on all {{parameters}}},\ }in\ \href {https://doi.org/10.1109/FOCS.2015.54} {\emph {\bibinfo {booktitle} {2015 {{IEEE}} 56th {{Annual Symposium}} on {{Foundations}} of {{Computer Science}}}}}\ (\bibinfo {year} {2015})\ pp.\ \bibinfo {pages} {792--809}\BibitemShut {NoStop}%
\bibitem [{\citenamefont {Berry}\ \emph {et~al.}(2014)\citenamefont {Berry}, \citenamefont {Childs}, \citenamefont {Cleve}, \citenamefont {Kothari},\ and\ \citenamefont {Somma}}]{berryExponentialImprovementPrecision2014}%
  \BibitemOpen
  \bibfield  {author} {\bibinfo {author} {\bibfnamefont {D.~W.}\ \bibnamefont {Berry}}, \bibinfo {author} {\bibfnamefont {A.~M.}\ \bibnamefont {Childs}}, \bibinfo {author} {\bibfnamefont {R.}~\bibnamefont {Cleve}}, \bibinfo {author} {\bibfnamefont {R.}~\bibnamefont {Kothari}},\ and\ \bibinfo {author} {\bibfnamefont {R.~D.}\ \bibnamefont {Somma}},\ }\bibfield  {title} {\bibinfo {title} {Exponential improvement in precision for simulating sparse hamiltonians},\ }in\ \href {https://doi.org/10.1145/2591796.2591854} {\emph {\bibinfo {booktitle} {Proceedings of the Forty-Sixth Annual ACM Symposium on Theory of Computing}}},\ \bibinfo {series and number} {STOC '14}\ (\bibinfo  {publisher} {Association for Computing Machinery},\ \bibinfo {address} {New York, NY, USA},\ \bibinfo {year} {2014})\ p.\ \bibinfo {pages} {283–292}\BibitemShut {NoStop}%
\bibitem [{\citenamefont {Berry}\ \emph {et~al.}(2015{\natexlab{b}})\citenamefont {Berry}, \citenamefont {Childs}, \citenamefont {Cleve}, \citenamefont {Kothari},\ and\ \citenamefont {Somma}}]{berrySimulatingHamiltonianDynamics2015b}%
  \BibitemOpen
  \bibfield  {author} {\bibinfo {author} {\bibfnamefont {D.~W.}\ \bibnamefont {Berry}}, \bibinfo {author} {\bibfnamefont {A.~M.}\ \bibnamefont {Childs}}, \bibinfo {author} {\bibfnamefont {R.}~\bibnamefont {Cleve}}, \bibinfo {author} {\bibfnamefont {R.}~\bibnamefont {Kothari}},\ and\ \bibinfo {author} {\bibfnamefont {R.~D.}\ \bibnamefont {Somma}},\ }\bibfield  {title} {\bibinfo {title} {Simulating {{Hamiltonian dynamics}} with a {{truncated Taylor series}}},\ }\href {https://doi.org/10.1103/PhysRevLett.114.090502} {\bibfield  {journal} {\bibinfo  {journal} {Phys. Rev. Lett.}\ }\textbf {\bibinfo {volume} {114}},\ \bibinfo {pages} {090502} (\bibinfo {year} {2015}{\natexlab{b}})}\BibitemShut {NoStop}%
\bibitem [{\citenamefont {Low}\ and\ \citenamefont {Chuang}(2019)}]{lowHamiltonianSimulationQubitization2019b}%
  \BibitemOpen
  \bibfield  {author} {\bibinfo {author} {\bibfnamefont {G.~H.}\ \bibnamefont {Low}}\ and\ \bibinfo {author} {\bibfnamefont {I.~L.}\ \bibnamefont {Chuang}},\ }\bibfield  {title} {\bibinfo {title} {Hamiltonian {{simulation}} by {{qubitization}}},\ }\href {https://doi.org/10.22331/q-2019-07-12-163} {\bibfield  {journal} {\bibinfo  {journal} {Quantum}\ }\textbf {\bibinfo {volume} {3}},\ \bibinfo {pages} {163} (\bibinfo {year} {2019})}\BibitemShut {NoStop}%
\bibitem [{\citenamefont {Wang}\ \emph {et~al.}(2024)\citenamefont {Wang}, \citenamefont {McArdle},\ and\ \citenamefont {Berta}}]{wangQubitEfficientRandomizedQuantum2024}%
  \BibitemOpen
  \bibfield  {author} {\bibinfo {author} {\bibfnamefont {S.}~\bibnamefont {Wang}}, \bibinfo {author} {\bibfnamefont {S.}~\bibnamefont {McArdle}},\ and\ \bibinfo {author} {\bibfnamefont {M.}~\bibnamefont {Berta}},\ }\bibfield  {title} {\bibinfo {title} {Qubit-{{efficient randomized quantum algorithms}} for {{linear algebra}}},\ }\href {https://doi.org/10.1103/PRXQuantum.5.020324} {\bibfield  {journal} {\bibinfo  {journal} {PRX Quantum}\ }\textbf {\bibinfo {volume} {5}},\ \bibinfo {pages} {020324} (\bibinfo {year} {2024})}\BibitemShut {NoStop}%
\bibitem [{\citenamefont {Wan}\ \emph {et~al.}(2022)\citenamefont {Wan}, \citenamefont {Berta},\ and\ \citenamefont {Campbell}}]{wanRandomizedQuantumAlgorithm2022}%
  \BibitemOpen
  \bibfield  {author} {\bibinfo {author} {\bibfnamefont {K.}~\bibnamefont {Wan}}, \bibinfo {author} {\bibfnamefont {M.}~\bibnamefont {Berta}},\ and\ \bibinfo {author} {\bibfnamefont {E.~T.}\ \bibnamefont {Campbell}},\ }\bibfield  {title} {\bibinfo {title} {Randomized {{quantum algorithm}} for {{statistical phase estimation}}},\ }\href {https://doi.org/10.1103/PhysRevLett.129.030503} {\bibfield  {journal} {\bibinfo  {journal} {Phys. Rev. Lett.}\ }\textbf {\bibinfo {volume} {129}},\ \bibinfo {pages} {030503} (\bibinfo {year} {2022})}\BibitemShut {NoStop}%
\bibitem [{\citenamefont {Vazquez}\ \emph {et~al.}(2023)\citenamefont {Vazquez}, \citenamefont {Egger}, \citenamefont {Ochsner},\ and\ \citenamefont {Woerner}}]{vazquezWellconditionedMultiproductFormulas2023}%
  \BibitemOpen
  \bibfield  {author} {\bibinfo {author} {\bibfnamefont {A.~C.}\ \bibnamefont {Vazquez}}, \bibinfo {author} {\bibfnamefont {D.~J.}\ \bibnamefont {Egger}}, \bibinfo {author} {\bibfnamefont {D.}~\bibnamefont {Ochsner}},\ and\ \bibinfo {author} {\bibfnamefont {S.}~\bibnamefont {Woerner}},\ }\bibfield  {title} {\bibinfo {title} {Well-conditioned multi-product formulas for hardware-friendly {{Hamiltonian}} simulation},\ }\href {https://doi.org/10.22331/q-2023-07-25-1067} {\bibfield  {journal} {\bibinfo  {journal} {Quantum}\ }\textbf {\bibinfo {volume} {7}},\ \bibinfo {pages} {1067} (\bibinfo {year} {2023})}\BibitemShut {NoStop}%
\bibitem [{\citenamefont {Osborne}(2006)}]{osborneEfficientApproximationDynamics2006}%
  \BibitemOpen
  \bibfield  {author} {\bibinfo {author} {\bibfnamefont {T.~J.}\ \bibnamefont {Osborne}},\ }\bibfield  {title} {\bibinfo {title} {Efficient {{approximation}} of the {{dynamics}} of {{one-dimensional quantum spin systems}}},\ }\href {https://doi.org/10.1103/PhysRevLett.97.157202} {\bibfield  {journal} {\bibinfo  {journal} {Phys. Rev. Lett.}\ }\textbf {\bibinfo {volume} {97}},\ \bibinfo {pages} {157202} (\bibinfo {year} {2006})}\BibitemShut {NoStop}%
\bibitem [{\citenamefont {Daley}\ \emph {et~al.}(2004)\citenamefont {Daley}, \citenamefont {Kollath}, \citenamefont {Schollw{\"o}ck},\ and\ \citenamefont {Vidal}}]{Daley}%
  \BibitemOpen
  \bibfield  {author} {\bibinfo {author} {\bibfnamefont {A.~J.}\ \bibnamefont {Daley}}, \bibinfo {author} {\bibfnamefont {C.}~\bibnamefont {Kollath}}, \bibinfo {author} {\bibfnamefont {U.}~\bibnamefont {Schollw{\"o}ck}},\ and\ \bibinfo {author} {\bibfnamefont {G.}~\bibnamefont {Vidal}},\ }\bibfield  {title} {\bibinfo {title} {{Time-dependent density-matrix renormalization-group using adaptive effective Hilbert spaces}},\ }\href {https://doi.org/10.1088/1742-5468/2004/04/P04005} {\bibfield  {journal} {\bibinfo  {journal} {J. Stat. Mech.}\ ,\ \bibinfo {pages} {P04005}} (\bibinfo {year} {2004})}\BibitemShut {NoStop}%
\bibitem [{\citenamefont {Or{\'u}s}(2019)}]{orusTensorNetworksComplex2019}%
  \BibitemOpen
  \bibfield  {author} {\bibinfo {author} {\bibfnamefont {R.}~\bibnamefont {Or{\'u}s}},\ }\bibfield  {title} {\bibinfo {title} {Tensor networks for complex quantum systems},\ }\href {https://doi.org/10.1038/s42254-019-0086-7} {\bibfield  {journal} {\bibinfo  {journal} {Nature Rev. Phys.}\ }\textbf {\bibinfo {volume} {1}},\ \bibinfo {pages} {538} (\bibinfo {year} {2019})}\BibitemShut {NoStop}%
\bibitem [{\citenamefont {Schuster}\ \emph {et~al.}(2024)\citenamefont {Schuster}, \citenamefont {Yin}, \citenamefont {Gao},\ and\ \citenamefont {Yao}}]{schusterPolynomialtimeClassicalAlgorithm2024}%
  \BibitemOpen
  \bibfield  {author} {\bibinfo {author} {\bibfnamefont {T.}~\bibnamefont {Schuster}}, \bibinfo {author} {\bibfnamefont {C.}~\bibnamefont {Yin}}, \bibinfo {author} {\bibfnamefont {X.}~\bibnamefont {Gao}},\ and\ \bibinfo {author} {\bibfnamefont {N.~Y.}\ \bibnamefont {Yao}},\ }\bibfield  {title} {\bibinfo {title} {A polynomial-time classical algorithm for noisy quantum circuits},\ }\href {http://doi.org/10.48550/arXiv.2407.12768} {\bibfield  {journal} {\bibinfo  {journal} {arXiv:2407.12768}\ } (\bibinfo {year} {2024})}\BibitemShut {NoStop}%
\bibitem [{\citenamefont {Wild}\ and\ \citenamefont {Alhambra}(2023)}]{wildClassicalSimulationShortTime2023}%
  \BibitemOpen
  \bibfield  {author} {\bibinfo {author} {\bibfnamefont {D.~S.}\ \bibnamefont {Wild}}\ and\ \bibinfo {author} {\bibfnamefont {{\'A}.~M.}\ \bibnamefont {Alhambra}},\ }\bibfield  {title} {\bibinfo {title} {Classical {{simulation}} of {{short-time quantum dynamics}}},\ }\href {https://doi.org/10.1103/PRXQuantum.4.020340} {\bibfield  {journal} {\bibinfo  {journal} {PRX Quantum}\ }\textbf {\bibinfo {volume} {4}},\ \bibinfo {pages} {020340} (\bibinfo {year} {2023})}\BibitemShut {NoStop}%
\bibitem [{\citenamefont {Shao}\ \emph {et~al.}(2024)\citenamefont {Shao}, \citenamefont {Wei}, \citenamefont {Cheng},\ and\ \citenamefont {Liu}}]{shaoSimulatingNoisyVariational2024}%
  \BibitemOpen
  \bibfield  {author} {\bibinfo {author} {\bibfnamefont {Y.}~\bibnamefont {Shao}}, \bibinfo {author} {\bibfnamefont {F.}~\bibnamefont {Wei}}, \bibinfo {author} {\bibfnamefont {S.}~\bibnamefont {Cheng}},\ and\ \bibinfo {author} {\bibfnamefont {Z.}~\bibnamefont {Liu}},\ }\bibfield  {title} {\bibinfo {title} {Simulating {{Noisy Variational Quantum Algorithms}}: {{A Polynomial Approach}}},\ }\href {https://doi.org/10.1103/PhysRevLett.133.120603} {\bibfield  {journal} {\bibinfo  {journal} {Physical Review Letters}\ }\textbf {\bibinfo {volume} {133}},\ \bibinfo {pages} {120603} (\bibinfo {year} {2024})}\BibitemShut {NoStop}%
\bibitem [{\citenamefont {Begu{\v s}i{\'c}}\ \emph {et~al.}(2024)\citenamefont {Begu{\v s}i{\'c}}, \citenamefont {Gray},\ and\ \citenamefont {Chan}}]{begusicFastConvergedClassical2024}%
  \BibitemOpen
  \bibfield  {author} {\bibinfo {author} {\bibfnamefont {T.}~\bibnamefont {Begu{\v s}i{\'c}}}, \bibinfo {author} {\bibfnamefont {J.}~\bibnamefont {Gray}},\ and\ \bibinfo {author} {\bibfnamefont {G.~K.-L.}\ \bibnamefont {Chan}},\ }\bibfield  {title} {\bibinfo {title} {Fast and converged classical simulations of evidence for the utility of quantum computing before fault tolerance},\ }\href {https://doi.org/10.1126/sciadv.adk4321} {\bibfield  {journal} {\bibinfo  {journal} {Science Advances}\ }\textbf {\bibinfo {volume} {10}},\ \bibinfo {pages} {eadk4321} (\bibinfo {year} {2024})}\BibitemShut {NoStop}%
\bibitem [{\citenamefont {Angrisani}\ \emph {et~al.}(2024)\citenamefont {Angrisani}, \citenamefont {Schmidhuber}, \citenamefont {Rudolph}, \citenamefont {Cerezo}, \citenamefont {Holmes},\ and\ \citenamefont {Huang}}]{angrisaniClassicallyEstimatingObservables2024}%
  \BibitemOpen
  \bibfield  {author} {\bibinfo {author} {\bibfnamefont {A.}~\bibnamefont {Angrisani}}, \bibinfo {author} {\bibfnamefont {A.}~\bibnamefont {Schmidhuber}}, \bibinfo {author} {\bibfnamefont {M.~S.}\ \bibnamefont {Rudolph}}, \bibinfo {author} {\bibfnamefont {M.}~\bibnamefont {Cerezo}}, \bibinfo {author} {\bibfnamefont {Z.}~\bibnamefont {Holmes}},\ and\ \bibinfo {author} {\bibfnamefont {H.-Y.}\ \bibnamefont {Huang}},\ }\bibfield  {title} {\bibinfo {title} {Classically estimating observables of noiseless quantum circuits},\ }\href {https://doi.org/10.48550/arXiv.2409.01706} {\bibfield  {journal} {\bibinfo  {journal} {arXiv:2409.01706}\ } (\bibinfo {year} {2024})}\BibitemShut {NoStop}%
\bibitem [{\citenamefont {Angrisani}\ \emph {et~al.}(2025)\citenamefont {Angrisani}, \citenamefont {Mele}, \citenamefont {Rudolph}, \citenamefont {Cerezo},\ and\ \citenamefont {Holmes}}]{angrisaniSimulatingQuantumCircuits2025a}%
  \BibitemOpen
  \bibfield  {author} {\bibinfo {author} {\bibfnamefont {A.}~\bibnamefont {Angrisani}}, \bibinfo {author} {\bibfnamefont {A.~A.}\ \bibnamefont {Mele}}, \bibinfo {author} {\bibfnamefont {M.~S.}\ \bibnamefont {Rudolph}}, \bibinfo {author} {\bibfnamefont {M.}~\bibnamefont {Cerezo}},\ and\ \bibinfo {author} {\bibfnamefont {Z.}~\bibnamefont {Holmes}},\ }\bibfield  {title} {\bibinfo {title} {Simulating quantum circuits with arbitrary local noise using {{Pauli Propagation}}},\ }\href {https://doi.org/10.48550/arXiv.2501.13101} {\bibfield  {journal} {\bibinfo  {journal} {arXiv:2501.13101}\ } (\bibinfo {year} {2025})}\BibitemShut {NoStop}%
\bibitem [{\citenamefont {Ippoliti}\ \emph {et~al.}(2021)\citenamefont {Ippoliti}, \citenamefont {Gullans}, \citenamefont {Gopalakrishnan}, \citenamefont {Huse},\ and\ \citenamefont {Khemani}}]{PhysRevX.11.011030}%
  \BibitemOpen
  \bibfield  {author} {\bibinfo {author} {\bibfnamefont {M.}~\bibnamefont {Ippoliti}}, \bibinfo {author} {\bibfnamefont {M.~J.}\ \bibnamefont {Gullans}}, \bibinfo {author} {\bibfnamefont {S.}~\bibnamefont {Gopalakrishnan}}, \bibinfo {author} {\bibfnamefont {D.~A.}\ \bibnamefont {Huse}},\ and\ \bibinfo {author} {\bibfnamefont {V.}~\bibnamefont {Khemani}},\ }\bibfield  {title} {\bibinfo {title} {Entanglement phase transitions in measurement-only dynamics},\ }\href {https://doi.org/10.1103/PhysRevX.11.011030} {\bibfield  {journal} {\bibinfo  {journal} {Phys. Rev. X}\ }\textbf {\bibinfo {volume} {11}},\ \bibinfo {pages} {011030} (\bibinfo {year} {2021})}\BibitemShut {NoStop}%
\bibitem [{\citenamefont {Skinner}\ \emph {et~al.}(2019)\citenamefont {Skinner}, \citenamefont {Ruhman},\ and\ \citenamefont {Nahum}}]{PhysRevX.9.031009}%
  \BibitemOpen
  \bibfield  {author} {\bibinfo {author} {\bibfnamefont {B.}~\bibnamefont {Skinner}}, \bibinfo {author} {\bibfnamefont {J.}~\bibnamefont {Ruhman}},\ and\ \bibinfo {author} {\bibfnamefont {A.}~\bibnamefont {Nahum}},\ }\bibfield  {title} {\bibinfo {title} {Measurement-induced phase transitions in the dynamics of entanglement},\ }\href {https://doi.org/10.1103/PhysRevX.9.031009} {\bibfield  {journal} {\bibinfo  {journal} {Phys. Rev. X}\ }\textbf {\bibinfo {volume} {9}},\ \bibinfo {pages} {031009} (\bibinfo {year} {2019})}\BibitemShut {NoStop}%
\bibitem [{\citenamefont {Childs}\ and\ \citenamefont {Wiebe}(2012)}]{childsHamiltonianSimulationUsing2012b}%
  \BibitemOpen
  \bibfield  {author} {\bibinfo {author} {\bibfnamefont {A.~M.}\ \bibnamefont {Childs}}\ and\ \bibinfo {author} {\bibfnamefont {N.}~\bibnamefont {Wiebe}},\ }\bibfield  {title} {\bibinfo {title} {Hamiltonian simulation using linear combinations of unitary operations},\ }\href {https://doi.org/10.26421/QIC12.11-12-1} {\bibfield  {journal} {\bibinfo  {journal} {Quant. Inf. Comp.}\ }\textbf {\bibinfo {volume} {12}},\ \bibinfo {pages} {901} (\bibinfo {year} {2012})}\BibitemShut {NoStop}%
\bibitem [{\citenamefont {Bakshi}\ \emph {et~al.}(2023)\citenamefont {Bakshi}, \citenamefont {Liu}, \citenamefont {Moitra},\ and\ \citenamefont {Tang}}]{bakshiLearningQuantumHamiltonians2023}%
  \BibitemOpen
  \bibfield  {author} {\bibinfo {author} {\bibfnamefont {A.}~\bibnamefont {Bakshi}}, \bibinfo {author} {\bibfnamefont {A.}~\bibnamefont {Liu}}, \bibinfo {author} {\bibfnamefont {A.}~\bibnamefont {Moitra}},\ and\ \bibinfo {author} {\bibfnamefont {E.}~\bibnamefont {Tang}},\ }\bibfield  {title} {\bibinfo {title} {Learning quantum {{Hamiltonians}} at any temperature in polynomial time},\ }\href {https://doi.org/10.48550/arXiv.2310.02243} {\bibfield  {journal} {\bibinfo  {journal} {arXiv:2310.02243}\ } (\bibinfo {year} {2023})}\BibitemShut {NoStop}%
\bibitem [{\citenamefont {Childs}\ \emph {et~al.}(2021)\citenamefont {Childs}, \citenamefont {Su}, \citenamefont {Tran}, \citenamefont {Wiebe},\ and\ \citenamefont {Zhu}}]{childsTheoryTrotterError2021}%
  \BibitemOpen
  \bibfield  {author} {\bibinfo {author} {\bibfnamefont {A.~M.}\ \bibnamefont {Childs}}, \bibinfo {author} {\bibfnamefont {Y.}~\bibnamefont {Su}}, \bibinfo {author} {\bibfnamefont {M.~C.}\ \bibnamefont {Tran}}, \bibinfo {author} {\bibfnamefont {N.}~\bibnamefont {Wiebe}},\ and\ \bibinfo {author} {\bibfnamefont {S.}~\bibnamefont {Zhu}},\ }\bibfield  {title} {\bibinfo {title} {Theory of {{Trotter error}} with {{commutator scaling}}},\ }\href {https://doi.org/10.1103/PhysRevX.11.011020} {\bibfield  {journal} {\bibinfo  {journal} {Phys. Rev. X}\ }\textbf {\bibinfo {volume} {11}},\ \bibinfo {pages} {011020} (\bibinfo {year} {2021})}\BibitemShut {NoStop}%
\bibitem [{\citenamefont {Ohliger}\ \emph {et~al.}(2013)\citenamefont {Ohliger}, \citenamefont {Nesme},\ and\ \citenamefont {Eisert}}]{Efficient}%
  \BibitemOpen
  \bibfield  {author} {\bibinfo {author} {\bibfnamefont {M.}~\bibnamefont {Ohliger}}, \bibinfo {author} {\bibfnamefont {V.}~\bibnamefont {Nesme}},\ and\ \bibinfo {author} {\bibfnamefont {J.}~\bibnamefont {Eisert}},\ }\bibfield  {title} {\bibinfo {title} {Efficient and feasible state tomography of quantum many-body systems},\ }\href {https://doi.org/10.1088/1367-2630/15/1/015024} {\bibfield  {journal} {\bibinfo  {journal} {New J. Phys.}\ }\textbf {\bibinfo {volume} {15}},\ \bibinfo {pages} {015024} (\bibinfo {year} {2013})}\BibitemShut {NoStop}%
\bibitem [{\citenamefont {Evans}\ \emph {et~al.}(2019)\citenamefont {Evans}, \citenamefont {Harper},\ and\ \citenamefont {Flammia}}]{evansScalableBayesianHamiltonian2019}%
  \BibitemOpen
  \bibfield  {author} {\bibinfo {author} {\bibfnamefont {T.~J.}\ \bibnamefont {Evans}}, \bibinfo {author} {\bibfnamefont {R.}~\bibnamefont {Harper}},\ and\ \bibinfo {author} {\bibfnamefont {S.~T.}\ \bibnamefont {Flammia}},\ }\bibfield  {title} {\bibinfo {title} {Scalable {Bayesian} {Hamiltonian} learning},\ }\href {https://doi.org/10.48550/arXiv.1912.07636} {\bibfield  {journal} {\bibinfo  {journal} {arXiv:1912.07636}\ } (\bibinfo {year} {2019})}\BibitemShut {NoStop}%
\bibitem [{\citenamefont {Bairey}\ \emph {et~al.}(2019)\citenamefont {Bairey}, \citenamefont {Arad},\ and\ \citenamefont {Lindner}}]{baireyLearningLocalHamiltonian2019}%
  \BibitemOpen
  \bibfield  {author} {\bibinfo {author} {\bibfnamefont {E.}~\bibnamefont {Bairey}}, \bibinfo {author} {\bibfnamefont {I.}~\bibnamefont {Arad}},\ and\ \bibinfo {author} {\bibfnamefont {N.~H.}\ \bibnamefont {Lindner}},\ }\bibfield  {title} {\bibinfo {title} {Learning a {{local Hamiltonian}} from {{local measurements}}},\ }\href {https://doi.org/10.1103/PhysRevLett.122.020504} {\bibfield  {journal} {\bibinfo  {journal} {Phys. Rev. Lett.}\ }\textbf {\bibinfo {volume} {122}},\ \bibinfo {pages} {020504} (\bibinfo {year} {2019})}\BibitemShut {NoStop}%
\bibitem [{\citenamefont {Kokail}\ \emph {et~al.}(2019)\citenamefont {Kokail}, \citenamefont {Maier}, \citenamefont {{van Bijnen}}, \citenamefont {Brydges}, \citenamefont {Joshi}, \citenamefont {Jurcevic}, \citenamefont {Muschik}, \citenamefont {Silvi}, \citenamefont {Blatt}, \citenamefont {Roos},\ and\ \citenamefont {Zoller}}]{kokailSelfverifyingVariationalQuantum2019}%
  \BibitemOpen
  \bibfield  {author} {\bibinfo {author} {\bibfnamefont {C.}~\bibnamefont {Kokail}}, \bibinfo {author} {\bibfnamefont {C.}~\bibnamefont {Maier}}, \bibinfo {author} {\bibfnamefont {R.}~\bibnamefont {{van Bijnen}}}, \bibinfo {author} {\bibfnamefont {T.}~\bibnamefont {Brydges}}, \bibinfo {author} {\bibfnamefont {M.~K.}\ \bibnamefont {Joshi}}, \bibinfo {author} {\bibfnamefont {P.}~\bibnamefont {Jurcevic}}, \bibinfo {author} {\bibfnamefont {C.~A.}\ \bibnamefont {Muschik}}, \bibinfo {author} {\bibfnamefont {P.}~\bibnamefont {Silvi}}, \bibinfo {author} {\bibfnamefont {R.}~\bibnamefont {Blatt}}, \bibinfo {author} {\bibfnamefont {C.~F.}\ \bibnamefont {Roos}},\ and\ \bibinfo {author} {\bibfnamefont {P.}~\bibnamefont {Zoller}},\ }\bibfield  {title} {\bibinfo {title} {Self-verifying variational quantum simulation of lattice models},\ }\href {https://doi.org/10.1038/s41586-019-1177-4} {\bibfield  {journal} {\bibinfo  {journal} {Nature}\ }\textbf {\bibinfo {volume} {569}},\ \bibinfo {pages} {355} (\bibinfo {year}
  {2019})}\BibitemShut {NoStop}%
\bibitem [{\citenamefont {Carrasco}\ \emph {et~al.}(2021)\citenamefont {Carrasco}, \citenamefont {Elben}, \citenamefont {Kokail}, \citenamefont {Kraus},\ and\ \citenamefont {Zoller}}]{carrascoTheoreticalExperimentalPerspectives2021}%
  \BibitemOpen
  \bibfield  {author} {\bibinfo {author} {\bibfnamefont {J.}~\bibnamefont {Carrasco}}, \bibinfo {author} {\bibfnamefont {A.}~\bibnamefont {Elben}}, \bibinfo {author} {\bibfnamefont {C.}~\bibnamefont {Kokail}}, \bibinfo {author} {\bibfnamefont {B.}~\bibnamefont {Kraus}},\ and\ \bibinfo {author} {\bibfnamefont {P.}~\bibnamefont {Zoller}},\ }\bibfield  {title} {\bibinfo {title} {Theoretical and {{experimental perspectives}} of {{quantum verification}}},\ }\href {https://doi.org/10.1103/PRXQuantum.2.010102} {\bibfield  {journal} {\bibinfo  {journal} {PRX Quantum}\ }\textbf {\bibinfo {volume} {2}},\ \bibinfo {pages} {010102} (\bibinfo {year} {2021})}\BibitemShut {NoStop}%
\bibitem [{\citenamefont {Hangleiter}\ \emph {et~al.}(2024)\citenamefont {Hangleiter}, \citenamefont {Roth}, \citenamefont {Eisert},\ and\ \citenamefont {Roushan}}]{HamiltonianLearning}%
  \BibitemOpen
  \bibfield  {author} {\bibinfo {author} {\bibfnamefont {D.}~\bibnamefont {Hangleiter}}, \bibinfo {author} {\bibfnamefont {I.}~\bibnamefont {Roth}}, \bibinfo {author} {\bibfnamefont {J.}~\bibnamefont {Eisert}},\ and\ \bibinfo {author} {\bibfnamefont {P.}~\bibnamefont {Roushan}},\ }\bibfield  {title} {\bibinfo {title} {{Precise Hamiltonian identification of a superconducting quantum processor}},\ }\href {https://doi.org/10.1038/s41467-024-52629-3} {\bibfield  {journal} {\bibinfo  {journal} {Nature Comm.}\ }\textbf {\bibinfo {volume} {15}},\ \bibinfo {pages} {9595} (\bibinfo {year} {2024})}\BibitemShut {NoStop}%
\bibitem [{\citenamefont {Kliesch}\ \emph {et~al.}(2019)\citenamefont {Kliesch}, \citenamefont {Kueng}, \citenamefont {Eisert},\ and\ \citenamefont {Gross}}]{klieschGuaranteedRecoveryQuantum2019}%
  \BibitemOpen
  \bibfield  {author} {\bibinfo {author} {\bibfnamefont {M.}~\bibnamefont {Kliesch}}, \bibinfo {author} {\bibfnamefont {R.}~\bibnamefont {Kueng}}, \bibinfo {author} {\bibfnamefont {J.}~\bibnamefont {Eisert}},\ and\ \bibinfo {author} {\bibfnamefont {D.}~\bibnamefont {Gross}},\ }\bibfield  {title} {\bibinfo {title} {Guaranteed recovery of quantum processes from few measurements},\ }\href {https://doi.org/10.22331/q-2019-08-12-171} {\bibfield  {journal} {\bibinfo  {journal} {Quantum}\ }\textbf {\bibinfo {volume} {3}},\ \bibinfo {pages} {171} (\bibinfo {year} {2019})}\BibitemShut {NoStop}%
\bibitem [{\citenamefont {Eisert}\ \emph {et~al.}(2020)\citenamefont {Eisert}, \citenamefont {Hangleiter}, \citenamefont {Walk}, \citenamefont {Roth}, \citenamefont {Markham}, \citenamefont {Parekh}, \citenamefont {Chabaud},\ and\ \citenamefont {Kashefi}}]{eisertQuantumCertificationBenchmarking2020a}%
  \BibitemOpen
  \bibfield  {author} {\bibinfo {author} {\bibfnamefont {J.}~\bibnamefont {Eisert}}, \bibinfo {author} {\bibfnamefont {D.}~\bibnamefont {Hangleiter}}, \bibinfo {author} {\bibfnamefont {N.}~\bibnamefont {Walk}}, \bibinfo {author} {\bibfnamefont {I.}~\bibnamefont {Roth}}, \bibinfo {author} {\bibfnamefont {D.}~\bibnamefont {Markham}}, \bibinfo {author} {\bibfnamefont {R.}~\bibnamefont {Parekh}}, \bibinfo {author} {\bibfnamefont {U.}~\bibnamefont {Chabaud}},\ and\ \bibinfo {author} {\bibfnamefont {E.}~\bibnamefont {Kashefi}},\ }\bibfield  {title} {\bibinfo {title} {Quantum certification and benchmarking},\ }\href {https://doi.org/10.1038/s42254-020-0186-4} {\bibfield  {journal} {\bibinfo  {journal} {Nature Rev. Phys.}\ }\textbf {\bibinfo {volume} {2}},\ \bibinfo {pages} {382} (\bibinfo {year} {2020})}\BibitemShut {NoStop}%
\bibitem [{\citenamefont {Kliesch}\ and\ \citenamefont {Roth}(2021)}]{klieschTheoryQuantumSystem2021a}%
  \BibitemOpen
  \bibfield  {author} {\bibinfo {author} {\bibfnamefont {M.}~\bibnamefont {Kliesch}}\ and\ \bibinfo {author} {\bibfnamefont {I.}~\bibnamefont {Roth}},\ }\bibfield  {title} {\bibinfo {title} {Theory of {{quantum system certification}}},\ }\href {https://doi.org/10.1103/PRXQuantum.2.010201} {\bibfield  {journal} {\bibinfo  {journal} {PRX Quantum}\ }\textbf {\bibinfo {volume} {2}},\ \bibinfo {pages} {010201} (\bibinfo {year} {2021})}\BibitemShut {NoStop}%
\bibitem [{\citenamefont {Brand{\~a}o}\ \emph {et~al.}(2022)\citenamefont {Brand{\~a}o}, \citenamefont {Kueng},\ and\ \citenamefont {Fran{\c c}a}}]{brandaoFasterQuantumClassical2022}%
  \BibitemOpen
  \bibfield  {author} {\bibinfo {author} {\bibfnamefont {F.~G. S.~L.}\ \bibnamefont {Brand{\~a}o}}, \bibinfo {author} {\bibfnamefont {R.}~\bibnamefont {Kueng}},\ and\ \bibinfo {author} {\bibfnamefont {D.~S.}\ \bibnamefont {Fran{\c c}a}},\ }\bibfield  {title} {\bibinfo {title} {Faster quantum and classical {{SDP}} approximations for quadratic binary optimization},\ }\href {https://doi.org/10.22331/q-2022-01-20-625} {\bibfield  {journal} {\bibinfo  {journal} {Quantum}\ }\textbf {\bibinfo {volume} {6}},\ \bibinfo {pages} {625} (\bibinfo {year} {2022})}\BibitemShut {NoStop}%
\bibitem [{\citenamefont {Bermejo}\ \emph {et~al.}(2024)\citenamefont {Bermejo}, \citenamefont {Braccia}, \citenamefont {Rudolph}, \citenamefont {Holmes}, \citenamefont {Cincio},\ and\ \citenamefont {Cerezo}}]{bermejoQuantumConvolutionalNeural2024}%
  \BibitemOpen
  \bibfield  {author} {\bibinfo {author} {\bibfnamefont {P.}~\bibnamefont {Bermejo}}, \bibinfo {author} {\bibfnamefont {P.}~\bibnamefont {Braccia}}, \bibinfo {author} {\bibfnamefont {M.~S.}\ \bibnamefont {Rudolph}}, \bibinfo {author} {\bibfnamefont {Z.}~\bibnamefont {Holmes}}, \bibinfo {author} {\bibfnamefont {L.}~\bibnamefont {Cincio}},\ and\ \bibinfo {author} {\bibfnamefont {M.}~\bibnamefont {Cerezo}},\ }\bibfield  {title} {\bibinfo {title} {Quantum {{Convolutional Neural Networks}} are ({{Effectively}}) {{Classically Simulable}}},\ }\href {https://doi.org/10.48550/arXiv.2408.12739} {\bibfield  {journal} {\bibinfo  {journal} {arXiv:2408.12739}\ } (\bibinfo {year} {2024})}\BibitemShut {NoStop}%
\bibitem [{\citenamefont {Fuller}\ \emph {et~al.}(2025)\citenamefont {Fuller}, \citenamefont {Tran}, \citenamefont {Lykov}, \citenamefont {Johnson}, \citenamefont {Rossmannek}, \citenamefont {Wei}, \citenamefont {He}, \citenamefont {Kim}, \citenamefont {Vu}, \citenamefont {Sharma}, \citenamefont {Alexeev}, \citenamefont {Kandala},\ and\ \citenamefont {Mezzacapo}}]{fullerImprovedQuantumComputation2025}%
  \BibitemOpen
  \bibfield  {author} {\bibinfo {author} {\bibfnamefont {B.}~\bibnamefont {Fuller}}, \bibinfo {author} {\bibfnamefont {M.~C.}\ \bibnamefont {Tran}}, \bibinfo {author} {\bibfnamefont {D.}~\bibnamefont {Lykov}}, \bibinfo {author} {\bibfnamefont {C.}~\bibnamefont {Johnson}}, \bibinfo {author} {\bibfnamefont {M.}~\bibnamefont {Rossmannek}}, \bibinfo {author} {\bibfnamefont {K.~X.}\ \bibnamefont {Wei}}, \bibinfo {author} {\bibfnamefont {A.}~\bibnamefont {He}}, \bibinfo {author} {\bibfnamefont {Y.}~\bibnamefont {Kim}}, \bibinfo {author} {\bibfnamefont {D.}~\bibnamefont {Vu}}, \bibinfo {author} {\bibfnamefont {K.}~\bibnamefont {Sharma}}, \bibinfo {author} {\bibfnamefont {Y.}~\bibnamefont {Alexeev}}, \bibinfo {author} {\bibfnamefont {A.}~\bibnamefont {Kandala}},\ and\ \bibinfo {author} {\bibfnamefont {A.}~\bibnamefont {Mezzacapo}},\ }\bibfield  {title} {\bibinfo {title} {Improved {{Quantum Computation}} using {{Operator Backpropagation}}},\ }\href {https://doi.org/10.48550/arXiv.2502.01897} {\bibfield  {journal}
  {\bibinfo  {journal} {arXiv:2502.01897}\ } (\bibinfo {year} {2025})}\BibitemShut {NoStop}%
\end{thebibliography}%
% \newpage
\appendix

\section{Tail bound for the truncated Taylor series}
\label{A:tail_bound}

This appendix provides details of arguments presented in the main text. Using insights from the ``Trotter error with 1-norm scaling'' lemma of Ref.~\cite[Lemma 1]{childsTheoryTrotterError2021}, we can bound the operator norm of the remainder of the truncated Taylor series of the time-evolution operator by using Taylor's remainder 
theorem with
\begin{equation}
    \Norm{\sum_{k=K+1}^\infty\frac{(-\ii Ht)^k}{k!}}\leq \frac{\left(\norm{H}t\right)^{K+1}}{(K+1)!},
\end{equation}
where we use the operator norm unless otherwise specified.
Note that this is different from a naive tail bound that would result in a factor of $\ee^{\norm{H}t}$ instead.
Furthermore, since
\begin{equation}
    (K+1)!\geq \frac{(K+1)^{K+1}}{\ee^K}=\ee\left(\frac{K+1}{\ee}\right)^{K+1},
\end{equation}
we find
\begin{equation}
    \frac{\left(\norm{H}t\right)^{K+1}}{(K+1)!}\leq \left(\frac{\norm{H}t\ee}{(K+1)}\right)^{K+1}\ee^{-1}\leq \epsilon
\end{equation}
as soon as
\begin{equation}
    K+1=\frac{\log(1/(\ee\epsilon))}{\mathrm{W}\left(\frac{1}{\norm{H}t}\log(1/(\ee\epsilon))\right)},
\end{equation}
where $\mathrm{W}$ denotes the Lambert $\mathrm{W}$ function defined as the inverse of $xe^x$.
Assuming $\norm{Ht}\leq 1$ yields a scaling of 
\begin{equation}
    K=\mathcal{O}\left(\frac{\log(1/\epsilon)}{\log(\log(1/\epsilon)}\right),
\end{equation}
recovering the standard bound from Ref.~\cite[Eq.~(4)]{berrySimulatingHamiltonianDynamics2015b}.
When using $r$ segments of time evolution operators $U(t/r)$ with 
$r\geq \norm{H}t$, the error required for each segment must be smaller than $\epsilon/r$, resulting in 
\begin{equation}
    K=\mathcal{O}\left(\frac{\log(r/\epsilon)}{\log(\log(r/\epsilon)}\right).
\end{equation}

\section{Direct Taylor expansion 
of \texorpdfstring{$O(t)$}{O(t)}}
\label{A:direct_expansion}
In Section~\ref{sec:expanding_dynamics}, we have constructed an approximation of $O(t)$ by 
concatenating two truncated Taylor series of the time evolution operator.
Consequently, the resulting approximation will also contain terms with higher powers of $t$ than the cutoff $K$. 
Although such an approximation will be more accurate, it differs from a direct Taylor expansion of the target function $F(H,t)=O(t)$.
We will now showcase such a direct approach for the example of time dynamics and the resulting trade-off in error and the number of terms in the linear combination of Pauli operators setting $r=1$ for simplicity.
\subsection{Expansion}
Instead of Eq.~\eqref{eq:o_concatenated}, we Taylor expand $O(t)$ directly as
\begin{align}
   \nonumber
    O(t)\!&=\!\sum_{k+k'=0 }^K\!\!\frac{\ii^{k-k'}t^{k+k'}}{k!k'!}H^kOH^{k'} \\
    &\label{eq:expansion}
    \approx\sum_{k+k'=0 }^K\sum_{\stackrel{\boldsymbol{m}\in\llbracket 1..L_H\rrbracket^k}{\boldsymbol{n}\in\llbracket 1..L_H\rrbracket^{k'}}} \gamma_{\boldsymbol{m},\boldsymbol{n}}(t)Q_{\boldsymbol{m},\boldsymbol{n}},
\end{align}
which again is a sum of $n$-qubit Pauli 
operators 
$Q_{\boldsymbol{m},\boldsymbol{n}}$ 
with 
corresponding 
coefficients 
$\gamma_{\boldsymbol{m},\boldsymbol{n}}(t)$, defined as
\begin{align}
    \label{eq:gammas}\gamma_{\boldsymbol{m},\boldsymbol{n}}(t)&=\frac{\ii^{k-k'}t^{k+k'}}{k!k'!}\alpha_{m_1}\cdots\alpha_{m_k}\alpha_{n_1}\cdots\alpha_{n_{k'}},\\
    Q_{\boldsymbol{m},\boldsymbol{n}}&=H_{m_1}\cdots H_{m_k} O H_{n_1}\cdots H_{n_{k'}}.
\end{align}
The number of terms in this linear combination of Pauli operators is upper-bounded by $\sum_{k+k'\leq K}L_H^{k+k'}=\mathcal{O}(L^K)$ and is thus quadratically smaller than when concatenating two Taylor approximations. As such, we cannot expect the same error of this approximation and must construct an appropriate bound on the systematic approximation error.

\subsection{Error bound}
We present a rigorous error bound.
To this end, we can use the ``Trotter error with 1-norm scaling'' lemma of Ref.~\cite{childsTheoryTrotterError2021}.
This theorem bounds the tail $\rest_K$ of product formulas above $K$-th order via
\begin{align}
        \Norm{\rest_K\left( \prod_{\ell=1}^{L_H} \ee^{\ii \alpha_\ell H_\ell t}\right)} 
      &= \Norm{\ee^{\ii H t}-\prod_{\ell=1}^{L_H} \ee^{\ii \alpha_\ell H_\ell t}}\\
      & = \frac{\left(\sum_{\ell=1}^{L_H} \Norm{\alpha_\ell \, H_\ell} t \right)^{K + 1 }}{(K+1)!}
      \nonumber
\end{align}
and can be adapted to find the error between the exact time evolution of some observable $O$ and the presented approximation, which can be bounded by
\begin{align}
    \Norm{\ee^{-\ii Ht}O\ee^{\ii Ht} - \widetilde{O}_K(t)}& \leq \Norm{\rest_{K}\left(\ee^{-\ii H t} O \ee^{\ii Ht}\right)}\\
    \nonumber
    &=\Norm{\rest_{K}\left(\ee^{-\ii H t} \ee^{\ln{O}}\ee^{\ii Ht}\right)}\\
    \nonumber
    &=\norm{O}\frac{\left(2\norm{H} t\right)^{K+1}}{(K+1)!},
    \nonumber
\end{align}
where we have lifted $O$ to fit the product formula form and recovered $\norm{O}$ since Ref.~\cite[Lemma 6]{childsTheoryTrotterError2021} has an additional factor of $\norm{\prod_{\ell=1}^{L_H} \ee^{\ii \alpha_\ell H_\ell t}}$ which vanishes for all anti-Unitary operators in the exponent but remains for $\ee^{\ln O}$.
The error is thus larger by a factor $2^{k+1}/3$ compared to the concatenated approach discussed in Section~\ref{subsec:systematic_error}.
Again, these results could be extended to a commutator bound using insights from Ref.~\cite{childsTheoryTrotterError2021}.
\end{document}